%Paper: patt-sol/9305007
%From: "Aric Hagberg" <hags@math.arizona.edu>
%Date: Tue, 18 May 1993 22:12:11 -0700 (MST)

\magnification=1200

\nopagenumbers
\headline={\ifnum\pageno>1\hss -- \tenrm\folio -- \hss
\else\hfil\fi}
\baselineskip 16pt
\centerline{\bf Pattern Formation in Dissipative
Nonvariational Systems:}
\centerline{\bf  The Effects of Front Bifurcations}
\vskip .3truein
\baselineskip 16pt
\centerline{Aric Hagberg}
\medskip
\centerline{\it Program in Applied Mathematics}
\centerline{\it University of Arizona}
\centerline{\it Tucson, AZ 85721}
\bigskip
\centerline{Ehud Meron}
\medskip
\centerline{\it Arizona Center for Mathematical Sciences}
\centerline{\it and Department of Mathematics}
\centerline{\it University of Arizona}
\centerline{\it Tucson, AZ 85721}
\vskip .5 truein
\baselineskip 15pt
\noindent
{\sl Abstract}
\medskip
Patterns in reaction-diffusion  systems often contain two spatial scales;
a long scale determined by a typical wavelength or domain size, and a
short scale pertaining to front structures separating different domains.
Such patterns naturally develop in bistable
and excitable systems, but may also appear far beyond Hopf and Turing
bifurcations. The global behavior of domain patterns strongly depends on
the fronts' inner structures. In this paper we study a symmetry breaking
front bifurcation expected to occur in a wide class of reaction-diffusion
systems, and the effects it has on pattern formation and
pattern dynamics. We extend previous works on this type of front bifurcation
and clarify the relations among them.
We show that the appearance of front multiplicity beyond the bifurcation
point allows the formation of persistent patterns rather than transient ones.
In a different parameter regime, we find that the front bifurcation outlines
a transition from oscillating (or breathing) patterns to traveling ones.
Near a boundary we find that fronts beyond
the bifurcation can reflect, while those below it
either bind to the  boundary or disappear.

\bigskip\noindent
%Short title: Front Bifurcations and Pattern Dynamics\hfil\break
%AMS classification scheme numbers:\hfil\break
%PACS numbers:  4265, 4720, 6460, 8220 \hfil

%\baselineskip 22pt
\vfil
\eject
\noindent
{\bf 1.~ Introduction}
\medskip
Pattern formation out of equilibrium can often be attributed
to the coupling of diffusion with local nonlinear dynamics. These
processes are explicit in reaction-diffusion systems such as chemical
reactions or electrical activation of biological membranes, and implicit
in systems whose large scale behavior
is governed by dissipative amplitude equations such as lasers.
The coupling of nonlinearity and
diffusion is most beautifully realized in two chemical systems; the
Belousov-Zhabotinsky (BZ) reaction, where a variety of traveling patterns
have been observed (for a recent review see [Mer92]), and the
Chlorite-Iodide-Malonic-Acid
(CIMA) reaction, where stationary, traveling, and turbulent patterns have
been found [CDB90,OuS91,LKE92].
Reaction-diffusion systems can be divided into three main categories:
Excitable, bistable, and Hopf-Turing systems. Patterns in
excitable and bistable systems normally involve front structures separating
domains of different uniform or quasi-uniform states. They coexist with one
or two stable uniform states and thus need to be triggered by
specific choices of initial conditions. A Hopf-Turing
system, on the other hand, gives rise to patterns through the destabilization
of a uniform stationary state in a Hopf or a Turing bifurcation [Tur52,Mur89].
Close to onset,
smooth oscillatory or stationary patterns usually appear. However, as the
system is driven away from onset, short scale structures, in the form of
boundary layers or fronts, may develop. In this regime, the patterns that
emerge resememble those in excitable and bistable systems.

In this paper we study patterns in reaction-diffusion systems consisting of
fronts separating different domains. Our primary concern here is with
the effect {\it front bifurcations} may have on the dynamics of fronts
and on the patterns they form. Most of our analysis will focus on bistable
systems, however, many of the results to be derived also apply to excitable
systems, and to Hopf-Turing systems far beyond onset.

We shall consider a specific reaction-diffusion model that exhibits the
same variety of patterns that have
been observed in both the BZ and CIMA reactions. It shares many
gross features with models of the BZ reaction [Fib85] and the Lengyel-Epstein
model of the CIMA reaction [LeE92], but is more amenable to analysis.
The model consists of
two scalar fields, $u(x,t)$ and $v(x,t)$ and contains four parameters;
the ratio, $\epsilon=T_u/T_v$, between the time scales associated with
the two fields, the ratio, $\delta=D_v/D_u$, between the two diffusion
constants, and two parameters, $a_1>0$ and $a_0$, characterizing the local
reaction dynamics. The model reads
%******************************  1.1  *************************************
$$u_t=u-u^3-v+u_{xx},\eqno(1.1a)$$
$$v_t=\epsilon(u-a_1v-a_0)+\delta v_{xx},\eqno(1.1b)$$
where the subscripts on $u$ and $v$ denote partial derivatives.
Models of this type have extensively been studied in the context of
excitable media where $\epsilon\ll 1$. The form with $\delta=0$
is known as the FitzHugh Nagumo model of nerve conduction [Fit61,NAY62].
In the following we will refer to (1.1) as the FitzHugh Nagumo (FHN) model for
$\delta\ne 0$ as well.

The stationary homogeneous states of (1.1) are determined by the intersection
points of the two nullclines $v=u-u^3$ and $v=(u-a_0)/a_1$. Three basic
cases, corresponding to the three types of reaction-diffusion systems
mentioned above, can be distinguished as Figs. 1 illustrate.
({\it i})~ The nullclines
intersect at a single point lying on one of the outer branches of the
cubic nullcline $v=u-u^3$ (Fig. 1$a$). ({\it ii})~ The nullclines
intersect at a single point lying on the middle branch of the cubic
nullcline (Fig. 1$b$). ({\it iii})~ The nullclines intersect at three points
each lying on a different branch of the cubic nullcline (Fig. 1$c$).
Note that these intersection points,
and consequently the stationary homogeneous states of (1.1),
are independent of the parameters $\epsilon$ and $\delta$. These parameters,
however, do affect the stability of the stationary homogeneous states, and
will serve in the following as bifurcation parameters. They also have physical
meanings that do not depend on the details of the FHN model.
Thus, with appropriate choices of $\epsilon$ and $\delta$, case  ({\it i})
can describe an excitable medium, case  ({\it ii}) a medium undergoing either
an Hopf or a Turing bifurcation, and case  ({\it iii}) a bistable medium.
Note also that when $a_0=0$, Eqs. (1.1) have an odd symmetry $(u,v)\to(-u,-v)$.
Analyzing the symmetric system will help us identifying symmetry breaking
bifurcations and multiplicity of stable front solutions.
Finally, a few words about notations. The solutions $u(x,t)$
and $v(x,t)$ of (1.1) obviously depend on the four parameters, $a_0$, $a_1$,
$\epsilon$, and $\delta$, but we will avoid displaying this dependence unless
we specifically wish to address it. In that case we will display only the
parameter(s) under consideration (see for example (3.25)). The same rule
holds for critical values of $\epsilon$ or $\delta$ at which bifurcations
occur (see (3.33)).

We begin in section 2 by first identifying Hopf and Turing bifurcations for
case ({\it ii}). We then use singular perturbation considerations to argue
that far beyond these instabilities front structures can develop, and that
these structures coincide, to leading order, with the single front solutions
of case ({\it iii}). This will allow, later on, extending some of the
conclusions we draw for bistable systems to Hopf-Turing systems driven far
beyond onset. In section 3
we proceed to case ({\it iii}) or bistable systems and
study a symmetry breaking front bifurcation that gives rise
to front multiplicity. This bifurcation is closely related to chiral
symmetry breaking in magnetic domain walls [LaN79] and in front structures
arising in forced oscillatory systems [CLH90]. Indeed, some of the results
to be derived here apply to these systems as well.
The apearance of front multiplicity has dramatic effects on the behavior
of patterns. For $\delta$ sufficiently small it allows for the formation
of persistent patterns (stable traveling waves) rather than transient ones,
whereas for $\delta$ large enough it is responsible for transitions from
oscillating or stationary domains to traveling ones. Another consequence
of the coexistence of multiple fronts is the possible reflection of
traveling fronts at boundaries. We study these
effects in section 4. In section 5 we describe the numerical procedures
we have used in our simulations. We conclude in section 6 with a brief
summary and a discussion of new problems motivated by the present work.

\bigskip\noindent
{\bf 2.~ The Development of Fronts Far Beyond Onset}
\medskip
The FHN model, like a number of other reaction-diffusion models [RoM92],
may undergo
both Turing and Hopf bifurcations. Figs. 2 show the patterns that
develop near and far from the onset of these instabilities. The appearance
of patterns involving two spatial scales far beyond onset can be
understood using singular perturbation theory as we now show.

For the sake of simplicity, consider the symmetric FHN model, that is
(1.1) with $a_0=0$, and assume further that $0<a_1<1$. These parameter
settings imply case ({\it ii}) (see Fig. 1$b$). The stationary
homogeneous state, $(u,v)=(0,0)$, loses stability in a Turing bifurcation
when
%********************************* 2.1 ***************************************
$$\mu^{-1}=\mu_{tu}^{-1}=2-a_1+2(1-a_1)^{1/2},~~~ {\rm and}~~~
\epsilon>1/a_1,\eqno(2.1)$$
where
$$\mu=\epsilon/\delta.$$
For $\mu^{-1}>\mu_{tu}^{-1}$ the stationary homogeneous state becomes
unstable to perturbations of finite wavenumbers.
%as the dispersion relations drawn in Fig. 3 show.
The same state loses stability in a Hopf bifurcation when
%********************************* 2.2  **************************************
$$\epsilon=\epsilon_H=1/a_1.\eqno(2.2)$$
For $\epsilon<\epsilon_H$ it becomes unstable to uniform
perturbations.

Far beyond the Turing bifurcation  $\mu=\epsilon/\delta\ll 1$.
Introducing a rescaled space coordinate $z=\sqrt\mu x$, we
find that stationary solutions of (1.1) should satisfy
%********************************* 2.3 **********************************
$$\mu u_{zz}+u-v-u^3=0,\eqno(2.3a)$$
$$v_{zz}+u-a_1 v=0.\eqno(2.3b)$$
In all regions where $u$ varies on the same scale as $v$ we can neglect
the second derivative term in (2.3$a$) and solve the remaining cubic
equation for $u$ in terms of $v$. There are three solution branches of
which the outer two, $u=u_-(v)$ and $u=u_+(v)$, represent, respectively,
low and high $u$ values. In the following we will call regions
of high and low $u$ values up state and down state domains, respectively.
Using  $u_\pm(v)$ in (2.3$b$) we obtain closed
equations for $v$ in the up and down state domains.
There also might be regions where the second derivative term
in $(2.3a)$ balances the remaining terms. In these regions $u$ varies on
a scale of order $\sqrt\mu$ which is much shorter than the scale of order
unity over which $v$ varies. Such boundary layers arise when nearby regions
converge to different branches $u_\pm(v)$ (because of appropriate initial
conditions). Stretching back the space coordinate we find to leading order:
%******************************* 2.4 *************************************
$$u_{xx}+u-v-u^3=0~~~~~~~~~~v_{xx}=0.\eqno(2.4)$$
These equations have two symmetric {\it front} solutions
%******************************* 2.5 *************************************
$$v=0,~~~~~~~~~~u=\pm\tanh(x/\sqrt 2).\eqno(2.5)$$
To construct approximate solutions representing periodic arrays
of front structures as in Fig.
2$b$, we solve the closed equations for $v$ in the up and down state domains,
$v_{zz}+u_\pm(v)-a_1 v=0$, and match the solutions and their first
derivatives at the front regions. For short wavelength patterns we may
simplify the equations for $v$ by linearizing $u_\pm(v)$ around $v=0$:
$u_\pm(v)\approx\pm 1-v/2$. Carrying out this calculation we find a band
of periodic stationary solutions with up and down state domains of equal
size, a consequence of the odd symmetry of the model when $a_0= 0$. For
more details about this type of calculation the reader is referred to
Refs. [DKT88,KeL89,Mer92] as well as to Appendix B, where the size of a
single domain structure in an asymmetric bistable system is calculated.

Similar considerations apply to the Hopf bifurcation case. Far beyond that
bifurcation $\epsilon\ll\epsilon_H$, or $\epsilon\ll 1$ if we choose
$a_1\sim {\cal O}(1)$. Traveling wave solutions of (1.1) satisfy
%********************************* 2.6 ************************************
$$\epsilon u_{\zeta\zeta}+\epsilon c_0 u_\zeta+u-v-u^3=0,\eqno(2.6a)$$
$$\delta v_{\zeta\zeta}+c_0 v_\zeta+u-a_1 v=0,
\eqno(2.6b)$$
where $\zeta:=\sqrt\epsilon(x-ct)$ and $c=c_0\sqrt\epsilon$ is the traveling
wave speed. This choice of $c$ is appropriate for short wavelength
patterns [DKT88], but other choices may work as well [Fif85].
As before, we distinguish between domains where
both $u$ and $v$ vary on a scale of order unity, and fronts where
$u$ varies on a scale much shorter than that of $v$, this time of
order $\sqrt\epsilon$. For domains we find again the relations,
$u=u_\pm(v)$, and obtain the closed equations for
$v$:~ $\delta v_{\zeta\zeta}+c_0 v_\zeta+u_\pm(v)-a_1 v=0$.
To study the fronts we stretch the moving coordinate frame,
$\chi=\zeta/\sqrt\epsilon=x-ct$, and find the leading order equations
%***************************** 2.7 *************************************
$$u_{\chi\chi}+cu_{\chi}+u-v-u^3=0~~~~~~~~~~
\delta v_{\chi\chi}+c v_{\chi}=0.\eqno(2.7)$$
These equations have the front solution
%***************************** 2.8 **************************************
$$v=v_f,~~~~~~~~~~u=[u_-(v_f) e^{q\chi}+u_+(v_f)]/[1+e^{q\chi}],
\eqno(2.8)$$
where $q=[u_+(v_f)-u_-(v_f)]/\sqrt 2$ and $v_f$ is the (approximately)
constant value of $v$ across the front. A symmetric front solution is
obtained from (2.8) by the transformation $x\to -x$. Periodic traveling
solutions as appear in Fig. 2$d$ can be constructed as before.

Note that the singular perturbation analysis sketched above does not depend
on the parameter $a_1$ in any way that would affect the
qualitative results (at least for patterns whose wavelengths are
sufficiently short). In fact, for the particular model considered here, the
front structures are strictly independent of $a_1$ (see (2.5,8)). We
therefore expect to find similar patterns in bistable systems for which
$a_1>1$.

\bigskip\noindent
{\bf 3.~ A Front Bifurcation}
\medskip
Despite the vast literature on  FHN type models very few studies addressed
other than small $\epsilon$ values. The significance of broadening the
range of $\epsilon$ is that there exists a critical value $\epsilon_c$
at which a front bifurcation occurs. This bifurcation
has important implications on pattern formation as
we will see in sections 4 and 5. Rinzel and Terman [RiT82], who studied
the FHN model with $\delta=0$, were the first  to
observe the creation of fronts at a critical $\epsilon$ value. More
recently, Ikeda {\it et al.} [IMN89] analyzed the regime
$\delta/\epsilon\gg 1$, and found a pitchfork front bifurcation.
In this section we extend these results in a number of ways. First, we study
the FHN model with $\delta=0$
and obtain explicit forms for the front
solutions and for the bifurcation diagram in the vicinity of the bifurcation
point. Then we consider nonzero $\delta$ values and evaluate a bifurcation
line, $\delta=\delta_c(\epsilon)$, in the
$\epsilon - \delta$ plane. Finally, we connect these results to a
recent work by Coullet {\it et al.} [CLH90] where a nonequilibrium analog of
the transition from Ising to Bloch walls in ferromagnets with weak
anisotropy [LaN79] has been found. This relation gives a broader perspective
to the body of works on patterns in FHN type models. Before embarking on the
front bifurcation analysis we discuss a few limiting cases of (1.1) which
motivate some aspects we wish to emphasize later on.

\medskip\noindent
{\bf 3.1~ The variational case: $\epsilon=0$}
\smallskip

The simplest front structures appearing in dissipative systems are
those involving a single scalar field. One example, known
in the context of equilibrium phase transitions, is the time dependent
Ginzburg-Landau (TDGL) model
obtained from (1.1) by either setting $\epsilon=0$, or regarding $v$ to be
a constant parameter:
%****************************** 3.1 ***************************************
$$u_t=u-u^3-v+u_{xx}.\eqno(3.1)$$
This equation has the variational form
%****************************** 3.2 ****************************************
$$u_t=-\delta{\cal F}/\delta u,\eqno(3.2)$$
where ${\cal F}$ is a Lyapunov functional given by
%****************************** 3.3 ****************************************
$${\cal F}=\int \big[{\cal E}(u,v)
+u_x^2/2\bigr]dx,~~~~~~~~{\cal E}(u,v)=-u^2/2+u^4/4+vu.
\eqno(3.3)$$
In the terminology of equilibrium phase transitions, $u$ is an order
parameter, $v$ is an external field, and ${\cal F}$ is a free energy.
For $v$ values in the range $-2/(3\sqrt 3)<v<2/(3\sqrt 3)$ the free energy
density, ${\cal E}$,  has double-well forms as shown in Figs. 3. The two wells
correspond to the stationary homogeneous states, $u_-(v)$  and $u_+(v)$,
that solve the cubic equation $u^3-u+v=0$.

Front solutions of (3.1) (or domain walls separating different phases)
propagate in a preferred direction dictated by the minimization of ${\cal F}$.
The speed of propagation, $c$, is
determined by the nonlinear eigenvalue problem
%*********************************  3.4  **********************************
$$u^{\prime\prime}+cu^\prime+u-v-u^3=0,\eqno(3.4a)$$
$$u(\chi)=\cases {u_+(v)&$\chi\to -\infty$\cr
u_-(v)&$\chi\to\infty$,\cr}\eqno(3.4b)$$
where $\chi=x-ct$. An expression for $c$ can be
obtained by multiplying $(3.4a)$ by $u^\prime$ and
integrating along the whole line. This yields
%******************************  3.5  **********************************
$$c=c(v)
%=\alpha\int_{\phi_-(h)}^{\phi_+(h)} f(\phi,h)d\phi
=\alpha(v)\Bigl[{\cal E}\bigl(u_-(v)\bigr)-{\cal E}\bigl(u_+(v)\bigr)\Bigr],
\eqno(3.5)$$
where $\alpha(v)=1/\int_{-\infty}^\infty {u^\prime(v)}^2 d\chi$ is
positive.
For $v=0$ the two wells of $~{\cal E}$ are of equal depth
and the front solution of (3.4) is stationary ($c=0$). For negative $v$
values, ${\cal E}(u_-)>{\cal E}(u_+)$ and the
front moves in the positive $x$ direction ($c>0$) so as to increase that
part of the system having lower energy. When $v$ is positive
the front propagates toward negative $x$ values  ($c<0$).
The explicit form of this front solution is given in (2.8). The symmetric
solution is obtained by the transformation $\chi\to -\chi$ and $c\to -c$.

\medskip\noindent
{\bf 3.2~ The regime $\epsilon\gg 1$}
\smallskip

We turn now to the full form of (1.1) with $\delta=0$, and study the case
of bistability ({\it iii}) for large $\epsilon$.
We denote the two stable, stationary homogeneous states
by $(u_\pm,v_\pm)$, where the subscripts $+(-)$ refer to high (low) $u$ and $v$
values (see Fig. 1$c$), and ask what types of fronts connect these two states.

In the regime of large $\epsilon$, the field $v$
varies on a time scale much shorter than that of $u$, at least initially.
Stretching the time coordinate by the factor $\epsilon$, we find
%**************************** 3.6 **************************************
$$u_\tau=\nu (u-u^3-v+u_{xx}),\eqno(3.6a)$$
$$v_\tau=u-a_1v-a_0,\eqno(3.6b)$$
where $\nu=\epsilon^{-1}\ll 1$ and $\tau=t/\nu$. On that new scale, the
leading order form of $u$ is independent of $\tau$ (set $\nu=0$ in (3.6$a$))
and equation (3.6$b$) can be solved for $v$:
%**************************** 3.7 **************************************
$$v(x,t)=v_0(x) e^{-\epsilon a_1t}+
a_1^{-1}[u(x,t)-a_0][1-e^{-\epsilon a_1t}].\eqno(3.7)$$
Equation (3.7) implies that apart from a short transient of duration
$t\sim {\cal O}(\epsilon^{-1})$, the $v$ field follows adiabatically the
$u$ field:~ $v=(u-a_0)/a_1$. Using this form in (3.6$a$) or (1.1$a$)
we obtain the asymptotic system
%****************************** 3.8 ************************************
$$u_t=(1-a_1^{-1})u-u^3-a_0a_1^{-1}+u_{xx},\eqno(3.8a)$$
$$v=(u-a_0)/a_1.\eqno(3.8b)$$
Equation (3.8$a$) has the same variational structure as (3.1) has, and
consequently, all the properties discussed in section 3.1 apply here
as well. In particular, the property of fronts propagating in the
direction dictated by the minimization of the free energy is retained.
Note that the original system (1.1) is not variational, but reduces to
a variational one in the large $\epsilon$ regime.

\medskip\noindent
{\bf 3.3~ The regime $0<\epsilon\ll 1$}
\smallskip

The regime of small $\epsilon$ has been studied extensively (see the reviews
[TyK88,Mer92]). The question we address here is
how front solutions connecting the two states $(u_\pm,v_\pm)$ differ
from those we found for $\epsilon\gg 1$.

Note first that for $\epsilon\ll 1$ the $v$ field remains approximately
constant on the length scale over which $u$ varies. Thus, in the narrow front
region equation (3.1) applies with $v=v_f$, the value of $v$ at the front
position. The front speed is then determined by $c=c(v_f)$ where $c(v)$ is
given by (3.5). Imagine now that we prepare the system in the down state
$(u_-,v_-)$ and perturb it locally at the left edge of the system so as to
induce a transition to the upper
branch $u=u_+(v)$. This transition occurs at $v=v_-$ and, as a result, a
front propagating to the right, $c=c(v_-)>0$, will be induced as
Fig. 4$a$ demonstrates. If, on the other hand, the system is prepared in
the up state $(u_+,v_+)$ and perturbed on the right edge so as to induce
a transition to the lower branch $u=u_-(v)$, a front propagating to the
left, $c=c(v_+)<0$, will be induced as shown in Fig. 4$b$. Once the system
is on a given branch it converges to the stationary homogeneous state
lying on that branch. The two fronts therefore connect the same asymptotic
states as $\chi\to\pm\infty$, but propagate in {\it opposite} directions.
This property of (1.1) reflects its nonvariational structure.
Multiplicity of front solutions in reaction-diffusion systems with multiple
time scales
was first observed by Ortoleva and Ross  [OrR75]. The stability of this type of
solutions has been proved later by Rinzel and
Terman [RiT82]. Note in this respect that the front solution of (3.8),
calculated for $\epsilon\gg 1$,
is independent of $\epsilon$ and therefore continues to exist for
$\epsilon\ll 1$ as well. In this regime, however, it is no longer stable.

The picture revealed so far is as follows: For $\epsilon\gg 1$
there exists only one front solution having the asymptotic behavior
$(u,v)\to (u_\pm,v_\pm)$ as $\chi\to\mp\infty$.
For $\epsilon\ll 1$ three such front solutions coexist, two
of which are stable. In the next section we study the intermediate $\epsilon$
regime where we expect to find a bifurcation from a single to multiple front
solutions.

\medskip\noindent
{\bf 3.4~ The front bifurcation}
\smallskip

We consider the symmetric case ($a_0=0$) for which front solutions
of (3.8) are stationary. In the following we will always assume the asymptotic
behavior, $(u,v)\to (u_\pm,v_\pm)$ as $\chi\to\mp\infty$, unless we
specifically mention otherwise. The explicit form of the stationary
front solution of (3.8) is (see (2.8))
%***************************** 3.9 **************************************
$$u_s(x)=-u_+\tanh(\eta x),~~~~~~~~~~v_s(x)=a_1^{-1}u_s(x),\eqno(3.9)$$
where $u_+=-u_-=\sqrt(1-a_1^{-1})$, and $\eta=u_+/\sqrt2$.
We assume that at some $\epsilon=\epsilon_c$ propagating front solutions,
$u=u_p(\chi),~~v=v_p(\chi)$, where $\chi=x-ct$,
bifurcate from (3.9). In the vicinity of the bifurcation point we can
expand a propagating front solution in powers of its speed $c$:
%****************************** 3.10 ***********************************
$$\eqalign{
u_p(\chi)&=u_s(\chi)+cu_1(\chi)+c^2u_2(\chi)+...\cr
v_p(\chi)&=v_s(\chi)+cv_1(\chi)+c^2v_2(\chi)+...,\cr}\eqno(3.10)$$
as well as $\epsilon$:
%****************************** 3.11 ***********************************
$$\epsilon=\epsilon_c+c\epsilon_1+c^2\epsilon_2+....\eqno(3.11)$$
We insert these expansions into the propagating front equations
%****************************** 3.12  *********************************
$$\eqalign
{u_p^{\prime\prime}+cu_p^\prime+u_p-v_p-u_p^3 =0\cr
cv_p^\prime+\epsilon(u_p-a_1v_p)=0,\cr}\eqno(3.12)$$
collect all terms that contribute to a given order, and solve the resulting
equations for the corrections $u_i$'s and $v_i$'s successively as described
below. Note that the corrections to all orders should decay to zero as
$\chi\to\pm\infty$ because the leading order form, that is the stationary
solution, already satisfies the correct asymptotic limits.
\smallskip\noindent
At order $c$ we obtain:\hfil\break
%****************************** 3.13 **************************************

$${\cal M}u_1=\Bigl({1\over \epsilon_c a_1^2}-1\Bigr)u_s^\prime,~~~~~~~~~~
v_1=a_1^{-1}u_1+u_s^\prime,\eqno(3.13)$$
where ${\cal M}$ is the linear operator
%******************************* 3.14 ************************************
$${\cal M}=d^2/d\chi^2+1-a_1^{-1}-3u_s^2.\eqno(3.14)$$
We consider a space of functions that decay to zero as $\chi\to\pm\infty$
(e.g. Schwartz space). On this space ${\cal M}$ is self adjoint and has
the null vector $u_s^\prime$:
%******************************** 3.15 ***********************************
$${\cal M}u_s^\prime={\cal M}^\dagger u_s^\prime=0.\eqno(3.15)$$
Solvability of (3.13) then yields
$\epsilon_c=a_1^{-2}$.
Using that result in (3.13$a$) we find ${\cal M}u_1=0$ and thus
$u_1=bu_s^\prime$,
where $b$ is arbitrary constant. Note, however, that
$u_s^\prime$ is a translation mode and that assigning a particular value to
$b$ defines the origin on the $\chi$ axis. For simplicity we choose $b=0$
and thus obtain
%******************************** 3.16 ************************************
$$u_1=0,~~~~~~~~~~v_1=u_s^\prime.\eqno(3.16)$$

\smallskip\noindent
At order $c^2$ we obtain:\hfil\break
%******************************** 3.17 ************************************
$${\cal M}u_2=-\epsilon_1 a_1^2 u_s^\prime + a_1 u_s^{\prime\prime},
{}~~~~~~~~~~v_2=a_1^{-1}u_2+a_1u_s^{\prime\prime}.\eqno(3.17)$$
Since $u_s$ is odd, solvability of (3.17) requires $\epsilon_1=0$.
Using that result in (3.17) we find
%******************************** 3.18 ***********************************
$$u_2={a_1\over 2}\chi u_s^\prime,~~~~~~~~~~v_2={1\over 2}\chi u_s^\prime
+a_1 u_s^{\prime\prime}.\eqno(3.18)$$
\smallskip\noindent
At order $c^3$ we obtain:\hfil\break
%********************************* 3.19 ***********************************
$${\cal M}u_3=a_1^2u_s^{\prime\prime\prime}-\epsilon_2 a_1^2 u_s^\prime,
\eqno(3.19)$$
which yields the solvability condition $\epsilon_2=-{2\over 5}(1-a_1^{-1})$.

Summing up these results we obtain
%******************************** 3.20 ************************************
$$\eqalign{
u_p(\chi)&=u_s+{1\over 2}c^2a_1\chi u_s^\prime+...\cr
v_p(\chi)&=a_1^{-1}u_s+cu_s^\prime+c^2\Bigl({1\over 2}\chi u_s^\prime
+a_1u_s^{\prime\prime}\Bigr)+...,\cr}\eqno(3.20)$$
and
%******************************** 3.21 ***********************************
$$\epsilon=\epsilon_c+\epsilon_2 c^2 +...,~~~~~~~~~~\epsilon_c=a_1^{-2}
{}~~~~~~~~\epsilon_2 =-2(1-a_1^{-1})/5.\eqno(3.21)$$

We verified these results by comparing the pitchfork bifurcation
implied by (3.21) with a numerically computed one. The comparison is shown in
Fig. 5$a$ and indicates good agreement up to speed values
$c\sim {\cal O}(10^{-1})$. Note that for the bistable case considered here
$a_1>1$ and consequently $\epsilon_2$ is negative. This ensures that the
bifurcation is supercritical.

It is instructive to look at the propagating solutions (3.20) up to corrections
of order $c$ (including). These can be written as
%****************************** 3.22 ************************************
$$u_p(\chi)\approx u_s(\chi),~~~~~~~~~~v_p(\chi)\approx a_1^{-1}u_s(\chi+ca_1).
\eqno(3.22)$$
Comparing with (3.9) we see that, to that order, the solutions for $u$ and
$v$ remain the same except that $v$ is {\it translated} with respect
to $u$ by amount proportional to the speed $c$. The direction of translation
determines the direction of propagation as Figs. 6 demonstrate. This
symmetry breaking is also reflected by the phase portraits of the front
solutions in the $u-v$ plane, as Fig. 7 shows. The stationary solution
is described by
a straight line, $v-a_1^{-1}u=0$, that does not break the odd symmetry of
(1.1) (with $a_0=0$), while the propagating solutions break the odd symmetry
and deviate from that line. This deviation is related
to the front speed through
%*************************************************************************
$$c=\alpha_s\int_{-\infty}^\infty(v_p-a_1^{-1}u_p)u_s^\prime d\chi,$$
where $\alpha_s=1/\int_{-\infty}^\infty(u_s^\prime)^2 d\chi$.

So far we have considered the symmetric model. In general the parameter
$a_0$ will not be zero, unless there exists an inherent symmetry in the
system that enforces that condition. When $a_0\ne 0$ the pitchfork front
bifurcation unfolds into a saddle-node bifurcation as the bifurcation
diagram shown in Fig. 5$b$ suggests. The saddle-node bifurcation
occurs at a critical value, $\epsilon_c(a_0)$, smaller than
$\epsilon_c=a_1^{-2}$.

\medskip\noindent
{\bf 3.5~ Ising vs. Bloch fronts}
\smallskip

An interesting distinction between the stationary and the propagating
front solutions in the symmetric model can be made once we define an angle
or phase in the $u-v$ plane, $\varphi=\arctan(v/u)$. Across the stationary
front the phase is constant except at the core where it suffers a jump by
$\pi$ and the norm $(u^2+v^2)^{1/2}$ vanishes. Across a propagating front,
on the other hand, the phase rotates smoothly by an angle $\pi$ and the norm
never vanishes. An analogous front bifurcation has been found recently by
Coullet {\it et al.} [CLH90] in the forced complex Ginzburg-Landau (CGL)
equation
%****************************** 3.23 ************************************
$$A_t=(\rho+i\theta)A+(1+i\alpha)A_{xx}-(1+i\beta)\vert A\vert^2A+\gamma\bar A,
\eqno(3.23)$$
where $A(x,t)$ is a complex field. This equation describes spatiotemporal
modulations of an oscillating medium, periodically forced at twice the
oscillation frequency with strength
$\gamma$. The effect of the forcing is to create two stable, phased
locked states whose phases, or $\arg(A)$ values, differ by $\pi$. The phase
$\arg(A)$ plays the same role as $\varphi$ does in the FHN model;
it remains constant or smoothly rotates across a front separating different
phase locked states for sufficiently large or small $\gamma$ values,
respectively. Coullet {\it et al.} [CLH90] proposed these types of fronts
solutions to be the
nonequilibrium analogs of Ising and Bloch walls in ferromagnets with weak
anisotropy (where $\arg(A)$ corresponds to the angle the magnetization
vector makes with an easy magnetization direction) [LaN79].
In the following we will occasionally follow this terminology and refer to the
two types of fronts in the FHN model as Ising and Bloch fronts.
%We note that Bloch walls in ferromagnets do not propagate because of the
%variational nature of this equilibrium system. However, when this variational
%structure is broken

\medskip\noindent
{\bf 3.6~ The bifurcation line in the $\epsilon-\delta$ plane}
\smallskip
The extension of the bifurcation analysis of section 3.4 to nonzero
$\delta$ values is rather tedious and we evaluate here the bifurcation
line only.
Even this calculation is not simple as the exact stationary front solution
is not known. We therefore consider the regime of small $\delta$ values
where the stationary front solution can be expanded in powers of $\delta$,
and evaluate a linear approximation for the bifurcation line valid for
$\delta\ll 1$. At the other extreme, $\delta/\epsilon\gg 1$, we use a singular
perturbation analysis, similar to that of Ikeda {\it et al.} [IMN89].
The complete bifurcation line is obtained numerically by
integrating (1.1) with Neumann boundary conditions.

Consider the symmetric model, that is, (1.1) with $a_0=0$. A stationary
front solution $u=u_s(x;\delta)~~v=v_s(x;\delta)$ satisfies
%******************************** 3.24 *******************************
$$u_s^{\prime\prime}+u_s-v_s-u_s^3=0,\eqno(3.24a)$$
$$\delta v_s^{\prime\prime}+\epsilon(u_s-a_1 v_s)=0.\eqno(3.24b)$$
Note that in these notations $u=u_s(x;0)$, $v=v_s(x;0)$ is the
stationary solution (3.9).
For $\delta\ll 1$ we can solve (3.24) perturbatively. Since we shall need the
form of the stationary solution on the bifurcation line we set
$\epsilon=\epsilon_c(\delta)$ in (3.24$b$). Expanding the stationary solution,
$u=u_s(x;\delta)~~v=v_s(x;\delta)$, and $\epsilon_c(\delta)$ in powers
of $\delta$,
%******************************** 3.25 ********************************
$$\eqalign{
u_s(x;\delta)=u_s(x;0)+\delta u_1(x)+...,\cr
v_s(x;\delta)=v_s(x;0)+\delta v_1(x)+...,\cr}\eqno(3.25)$$
%******************************** 3.26 *********************************
$$\epsilon_c(\delta)=a_1^{-2}+\delta\epsilon_{c1}+...,\eqno(3.26)$$
and using these expansions in (3.24) we find at order $\delta$
%******************************** 3.27 ********************************
$${\cal M}u_1(x)=u_s^{\prime\prime}(x;0),\eqno(3.27)$$
%******************************** 3.28 *********************************
$$v_1(x)=a_1^{-1}u_1(x)+u_s^{\prime\prime}(x;0),\eqno(3.28)$$
where ${\cal M}$ is given by (3.14) with $u_s=u_s(x;0)$.
The solution of (3.27) is $u_1(x)={1\over 2}xu_s^\prime(x;0)$.
This, together with (3.28), yields the leading order approximation
%******************************** 3.29 *******************************
$$u_s(x;\delta)=u_s(x;0)+{1\over 2}\delta x u_s^\prime(x;0)+...,
\eqno(3.29a)$$
$$v_s(x;\delta)=a_1^{-1}u_s(x;0)+{1\over 2}\delta a_1^{-1}x u_s^\prime(x;0)
+\delta u_s^{\prime\prime}(x;0)+...,\eqno(3.29b)$$
where $u_s(x;0)$ is given by (3.9).

As before we expand $\epsilon$ and the propagation front solutions
in power series in $c$ (see (3.10))
%********************************* 3.30 ******************************
$$\eqalign{
u_p(\chi;\delta)&=u_s(\chi;\delta)+cu_1(\chi;\delta)+...,\cr
v_p(\chi;\delta)&=v_s(\chi;\delta)+cv_1(\chi;\delta)+...,\cr}\eqno(3.30)$$
%****************************** 3.31 ***********************************
$$\epsilon=\epsilon_c(\delta)+c\epsilon_1(\delta)+...,\eqno(3.31)$$
and insert these expansions into (1.1). At order $c$ we obtain
%****************************** 3.32 ************************************
$$\left(\matrix{\partial_\chi^2+1-3u_s^2(\chi;\delta)&-1\cr
\epsilon_c/\delta&\partial_\chi^2-\epsilon_ca_1/\delta\cr}\right)
\left(\matrix{u_1(\chi;\delta)\cr
v_1(\chi;\delta)\cr}\right)=~~~~~~~~~~$$
$$~~~~~~~~~~~~~~~-\left(\matrix{u_s^\prime(\chi;\delta)\cr
v_s^\prime(\chi;\delta)/\delta
+\epsilon_1[u_s(\chi;\delta)-a_1v_s(\chi;\delta)]/\delta\cr}\right).
\eqno(3.32)$$
Solvability of (3.32) requires the right hand side of this equation
to be orthogonal to the vector $(u_s^\prime(\chi;\delta),
-{\delta\over\epsilon_c}v_s^\prime(\chi;\delta))^T$. This yields the
following equation for the bifurcation line:
%**************************** 3.33 ********************************
$$\epsilon=\epsilon_c(\delta)=
\int_{-\infty}^\infty v_s^\prime(\chi;\delta)^2d\chi
\Bigl/\int_{-\infty}^\infty u_s^\prime(\chi;\delta)^2d\chi\Bigr..
\eqno(3.33)$$
Using the leading order approximation (3.29) in (3.33) we find
%****************************** 3.34 **********************************
$$\epsilon_c(\delta)=(1-b\delta)/a_1^2~~~~~{\rm or}~~~~~
\delta_c(\epsilon)=(1-a_1^2\epsilon)/b~~~~~~~~(\delta\ll 1),\eqno(3.34)$$
where $b=4(a_1-1)/5$.

In Appendix A we evaluate the bifurcation line for $\delta/\epsilon\gg 1$
using singular perturbation theory. This approach differs from that
presented above in that it uses the smallness of $\epsilon/\delta$ rather
than that of the speed. As noted by Ikeda {\it et al.} [IMN89] it yields
approximate
front solutions and bifurcation diagrams valid away from the bifurcation point
as well. The price though is the limited validity range along the
bifurcation line. Using this approach we obtain
%****************************** 3.35 ************************************
$$\epsilon_c(\delta)=d/\delta,~~~~~~~~~~d=9/(2a_1+1)^3,\eqno(3.35)$$
where we took $a_0=0$ (symmetric model).

Fig. 8 shows the exact bifurcation line obtained by direct
numerical integration of (1.1).
The linear approximation for $\delta\ll 1$ and the singular perturbation
result (3.35) (valid for $\delta/\epsilon\gg 1$) are shown in solid and
dashed lines, respectively.
We conclude this section by showing in Figs. 9 typical phase portraits of the
front solutions (more precisely, projections thereof on the $(u, v)$ plane)
for $\delta$ values of order unity or larger. The outer portraits pertain to
the two counter
propagating front solutions that exist for $\epsilon<\epsilon_c(\delta)$,
whereas the middle one represents the stationary front solution at
$\epsilon>\epsilon_c(\delta)$. Note that the field $v$ barely changes across
the front. This is a consequence of the smallness of $\epsilon/\delta$
(compare with the corresponding phase portraits  shown in
Fig. 7 for $\delta=0$).

\bigskip\noindent
{\bf 4.~ Implications on Pattern Formation}
\medskip\noindent
{\bf 4.1~ The Emergence of Persistent Patterns}
\smallskip
A significant implication of the front bifurcation for $\delta=0$ or
sufficiently small is that
it tells us where in parameter space we should expect an initial pattern of
domains to decay toward a uniform state and where to develop into a stable
traveling wave. Consider first the regime of high $\epsilon$ values,
$\epsilon\gg\epsilon_c(a_0)$ (see last paragraph of
section 3.4). In this parameter regime the nonvariational system (1.1)
reduces to the variational system (3.8). As a result, patterns evolve so
as to minimize the appropriate free energy functional.
Like in equilibrium phase transitions, up state domains either shrink or
expand (depending on the sign of $a_0$) to form the uniform state of lowest
free energy. The relaxation toward a uniform state occurs even for the
symmetric case ($a_0=0$) where an isolated front is stationary. As shown by
Kawasaki and Ohta [KaO82] and more recently by Carr and Pego [CaP89],
nearby fronts
that bound a domain attract and annihilate one another on a time
scale $t\sim\exp(\lambda/w)$ where $\lambda$ is the domain size and $w$ is
the front width. Thus, domains shrink and disappear, but the relaxation
can be extremely slow and unnoticible in practice.

We have found numerically that this qualitative behavior remains unchanged
for {\it all} $\epsilon>\epsilon_c(a_0)$. Fig. 10$a$ shows the relaxation
of an initial pattern toward a uniform state for
$\epsilon/\epsilon_c(a_0)=1.12$.
Fig. 10$b$ shows the time evolution of the same initial pattern for
$\epsilon/\epsilon_c(a_0)=.72$, that is beyond the bifurcation from Ising to
Bloch fronts. In contrast to Fig. 10$a$, here, a stable traveling pattern
develops. We explored the transition from transient to persistent patterns more
carefully by integrating equations (1.1) for various $\epsilon$ and $a_0$
values keeping the other parameters constant ($\delta=0$, $a_1=2$). The
resulting phase diagram is shown in Fig. 11. The solid (right) curve
designates the front bifurcation, $\epsilon=\epsilon_c(a_0)$. To the right
of this curve only one front solution $(u,v)\to(u_\pm,v_\pm)~ {\rm as}~
x\to\mp\infty$ exists, and initial patterns decay toward a uniform state.
To the left of this curve, three front solutions connecting the same
asymptotic states coexist; two stable (Bloch fronts) and one unstable (Ising
front). Stable traveling waves, however, appear only beyond a second threshold,
$\epsilon_p(a_0)$, given by the dashed (left) curve. Using a numerical
continuation method we found that for the symmetric case ($a_0=0$) traveling
wave solutions do exist for $\epsilon_p<\epsilon<\epsilon_c$, but they are
unstable. We did not find such solutions for the nonsymmetric model (i.e. for
$\epsilon_p(a_0)<\epsilon<\epsilon_c(a_0)$ with $a_0\ne 0$).

This dramatic change in the qualitative behavior of patterns can be
attributed to two related factors; the appearance of front multiplicity
and the appearance of a second independent field.
The former makes it possible for domains to travel rather than shrink or
expand, whereas the latter affects front interactions so as to bind a
trailing front to a leading one. In the rest of this section
we elaborate on these two factors in some length.

The multiplicity of front solutions for $\epsilon<\epsilon_c(a_0)$ and
the symmetry $x\to -x$ of (1.1) imply that
along with a front that transforms the lower state
$(u_-,v_-)$ to the upper state $(u_+,v_+$), there exists another
front (hereafter ``back'') propagating in the {\it same direction} that
transforms the upper state back to the lower one (note that the front and
the back represent two different Bloch fronts). A combination of the two,
using appropriate initial conditions, can yield a traveling up-state domain.
To be concrete, let us assume that
the (isolated) trailing front, or back, is faster than the leading front.
This situation is attenable with $a_0<0$. The fate of the traveling domain
as the back approaches the (leading) front depends on the interaction between
the two. Unlike the
case of Ising fronts where the field $v(x,t)$ is eliminated through the
relation $v=(u-a_0)/a_1$ and the interaction is attractive, for
$\epsilon<\epsilon_c(a_0)$, $v(x,t)$ is independent of $u(x,t)$ and allows for
a repulsive interaction as well.

To see this, let us assume that $\epsilon\ll\epsilon_p$ where
$\epsilon_p<\epsilon_c\sim{\cal O}(a_1^{-2})$.
%$a_1$ is sufficiently large so that
%$\epsilon<\epsilon_c(a_0)\ll 1$ (recall that $\epsilon_c(0)=a_1^{-2}$ and
%that $\epsilon_c(a_0)<\epsilon_c(0)$ for $a_0\ne 0$).
Then $v$ varies on time and length scales
much longer than those of $u$. Just behind the front the value of $v$ is still
close to the down state value $v_-$. It approches the up state value, $v_+$,
only far behind the front. Since $v$ is approximately constant across
the narrow back region, the back speed is determined by the local $v$ value,
$v_b$, according to (3.5). When the back is still far behind the front
$v_b\approx v_+$, the energy difference
${\cal V}\bigl(u_-(v_+)\bigr)-{\cal E}\bigl(u_+(v_+)\bigr)$ is large
(see Fig. 3$c$) and the
speed high. As it approaches the front, however, $v_b$ decreases, the energy
difference becomes smaller and the speed lower. Thus, the back is slowed down
(or repelled) by the field $v$ induced by the front. For $\epsilon$
sufficiently small the back approaches the front speed while it is still
far behind the front. At such distances the $u$ field follows adiabatically
the slow field $v$ (that is, $u=u_+(v)$) and a stable traveling domain of
fixed size is formed (see below).
For $\epsilon$ values closer to $\epsilon_c$ the adiabatic elimination of $u$
may not be valid and more complicated interactions are possible.
We believe that in this parameter range both repulsive
and attractive interactions are significant and suggest that the balance
between the two gives the threshold $\epsilon_p$ represented in Fig. 11 by the
inner curve. We note though that we have not studied yet this regime
in detail and that the above suggestion is not founded yet.

To study the formation of a stable traveling domain of fixed size we
assume $\epsilon\ll \epsilon_p$ and use a singular perturbation approach.
We first evaluate the slow field $v$ that builds up behind a front
propagating at constant speed $c=c(v_-)$. In this region $u=u_+(v)$ and the
equation we need to solve is
%****************************** 4.1 **********************************
$$cv^\prime+\epsilon\bigl(u_+(v)-a_1v-a_0\bigr)=0,\eqno(4.1)$$
where $v=v(\chi)$ and $\chi=x-ct$. The boundary conditions are
%****************************** 4.2 ***********************************
$$v(\chi_f^0)=v_-,~~~~~~~~~~~~v(\chi)\to v_+ ~~{\rm as }~~ \chi\to -\infty,
\eqno(4.2)$$
where $\chi_f^0$ is the front position in a frame moving at speed $c$.
To simplify (4.1) we take $a_1$ to be sufficiently large
so that $\vert v_-\vert\sim v_+\ll 1$. The branches $u_\pm$ can then be
linearized:
%****************************** 4.3 ***********************************
$$u_\pm(v)=\pm 1-v/2.\eqno(4.3)$$
Using (4.3) in (4.1) and solving for $v$ we obtain
%****************************** 4.4 *************************************
$$v(\chi)=(v_--v_+)e^{\epsilon\kappa(\chi-\chi_f^0)}+v_+,~~~~~~~~
\chi\le\chi_f^0,\eqno(4.4a)$$
where
$$v_\pm={\pm 1-a_0\over a_1+1/2},~~~~~~~~~~~~\kappa={a_1+1/2\over c}.
\eqno(4.4b)$$

A traveling domain of fixed size can be obtained by demanding the front
and back speeds to be equal
%****************************** 4.5 *************************************
$$-c(v_b^0)=c(v_-)=c.\eqno(4.5)$$
This relation together with (3.5) can be used to evaluate $v_b^0=v(\chi_b^0)$,
the level of $v$ at the back position for a domain of fixed size.
The fixed size of an up-state domain can be expressed in terms of $v_b^0$
using (4.4$a$):
%****************************** 4.6 *************************************
$$\lambda=\chi_f^0-\chi_b^0={1\over \epsilon\kappa}
\ln\Bigl({v_+-v_-\over v_+-v_b^0}\Bigr).\eqno(4.6)$$
Note that (4.6) requires $v_b^0<v_+$ or an isolated back to propagate faster
than an isolated front. For $v_b^0>v_+$ (back slower than front) the $v$ field
behind the front can never reach the value $v_b^0$ needed for a back to
propagate at the front speed. In this case, an up-state domain expands
indefinitely, and a down-state domain of fixed size (referred to as an E-pulse
in [RiT82]) becomes feasable. We illustrate these two cases in Figs. 12.

The relaxation toward an up-state domain of fixed size can be derived by
writing a back solution in the form
%****************************** 4.7 **************************************
$$u_b(x,t)=u_b^0(\chi-\chi_b)+\epsilon u_b^1(\chi-\chi_b,\epsilon t),
\eqno(4.7a)$$
$$v_b=v(\chi_b)=v_b^0+\epsilon v_b^1(\epsilon t),\eqno(4.7b)$$
where $u_b^0(\chi)$ solves
%****************************** 4.8 **************************************
$${u_b^0}^{\prime\prime}+c{u_b^0}^\prime+u_b^0-v_b^0-{u_b^0}^3=0,\eqno(4.8a)$$
$$u_b^0(\chi)=\cases {u_-(v_b^0)&$\chi\to -\infty$\cr
u_+(v_b^0)&$\chi\to\infty$,\cr}\eqno(4.8b)$$
and
%******************************    **************************************
$$\chi_b(\epsilon t)=\chi_b^0+\tilde\chi_b(\epsilon t)$$
is the actual back position. The slow time dependence of $v_b$ is introduced
to account for the declining $v$ field at the back position as the latter
approaches the front. Using (4.7) in (1.1$a$) we find
%****************************** 4.9 ***************************************
$$\partial_\chi^2 u_b^1+c\partial_\chi u_b^1 + (1-3{u_b^0}^2)u_b^1=v_b^1
-\epsilon^{-1}\dot\chi_b{du_b^0\over d\chi},\eqno(4.9)$$
where the dot over $\chi_b$ denotes differentiation with respect to the fast
time $t$. Solvability of (4.9) requires the right hand side of to be orthogonal
to $\exp(c\chi)du_b^0/d\chi$ or
%****************************** 4.10 ****************************************
$$\dot\chi_b=\beta\epsilon v_b^1,\eqno(4.10)$$
where
%******************************   *******************************************
$$\beta=\int_{-\infty}^\infty{du_b^0\over d\chi} e^{c\chi}d\chi\Bigl/
\int_{-\infty}^{\infty}\Bigl({du_b^0\over d\chi}\Bigr)^2 e^{c\chi}d\chi\Bigr.$$
is a positive constant.

Equation (4.10) relates the back speed to the level of $v$ at the back
position. The latter, in turn, can be related to former
through (4.4$a$):
%******************************* 4.11 ****************************************
$$\epsilon v_b^1=v(\chi_b)-v_b^0=(v_+-v_b^0)
\bigl(1-e^{\epsilon\kappa\tilde\chi_b}\bigr).\eqno(4.11)$$
Combining (4.10) and (4.11) we obtain an equation of motion for the back:
%******************************* 4.12 ***************************************
$$\dot{\tilde\chi_b}=\beta(v_+ - v_b^0)
\bigl(1-e^{\epsilon\kappa\tilde\chi_b}\bigr).\eqno(4.12)$$
The linearization of (4.12) about $\tilde\chi_b=0$ (i.e. the fixed size
domain) gives the equation
%*******************************    **************************************
$$\dot{\tilde\chi_b}=-\epsilon\kappa\beta(v_+-v_b^0)\tilde\chi_b.$$
Recall that $\epsilon$, $\kappa$ and $\beta$ are all positive and $v_b^0<v_+$
for a relaxation toward an up-state domain. We therfore conclude that the
up-state domain whose size is given by (4.6) is stable to translational
perturbations. Note that in the small $\epsilon$ regime considered here
the size of a domain is much larger than the front (back) width. Thus, the
front and back that bound a domain are approximately isolated (see (4.7$a$).
The isolated front solutions, in turn, are marginally stable to translations
but stable to other types of perturbations (such as amplitude modulations).
We therefore expect the
translational perturbations to be the most ``dangerous'' ones.

A similar approach can be used to study the relaxation of arrays of
traveling up-state domains toward periodic traveling waves. The difference
with respect to a single domain is that now the leading front of any domain
propagates into the slowly decaying $v$ field behind the back ahead of it.
As a result, the front speed is not constant; it rather depends on the
distance between the front and the leading back. A different kinematical
approach, leading to equations of motion expressed in implicit form, has
been proposed earlier by Keener [Kee80] in the context of excitable media.
We also note that the singular perturbation theory sketched above cannot
capture an oscillatory decay toward either of the stationary homogeneous
states. Such a decay, can lead to spatial chaos as discussed in refs.
[EMS88,Mer92].

So far we have considered the case $\delta=0$. Similar conclusions hold
for small enough $\delta$ values.
We conclude this section with a numerical study of the forced CGL equation
(3.23) for which $\delta=1$. We have added to the right hand side of this
equation a constant term, $\gamma_0$, to break
the symmetry between the two phase locked states (analogous to $a_0$). Such
a term appears when
the periodic forcing contains a component at the medium's oscillation
frequency. We have not explored in detail the ($\gamma,\gamma_0$) parameter
plane, but for fixed $\gamma_0$ we did observe transient patterns for $\gamma$
sufficiently large, stable traveling waves for $\gamma$ sufficiently small,
and transient patterns involving traveling domains in a narrow range of
intermediate $\gamma$ values. This suggests that the qualitative form of
the phase diagram drawn in Fig. 11 is quite general and might apply to
any system (with $\delta$ relatively small) undergoing an Ising-Bloch type
front bifurcation.

\medskip\noindent
{\bf 4.2~ Stationary, Oscillating and Traveling Domains in an Infinite System}
\smallskip
When the diffusion of $v$ is fast enough stable stationary domains, rather
than transient ones, may develop [EHT84] in the regime of Ising fronts.
The formation of this type of domain for $\delta/\epsilon$ sufficiently large
is illustrated in Fig. 13. As the two fronts approach
one another the $v$ field, inside the domain bounded by the fronts,
developes a narrow lumped structure.
The consequent diffusive damping lowers the level of $v$ at the front
regions and thus slow them down. The size, $\lambda$, of a
single stationary domain is given by
%******************************* 4.13 **********************************
$$\lambda=\varpi(\delta/\epsilon)^{1/2},~~~~~~~~~~
\varpi=-\ln(-a_0)/(a_1+{1\over 2})^{1/2},
\eqno(4.13)$$
assuming $\mu=\epsilon/\delta\ll 1$. We derive this result
in Appendix B using a singular perturbation approach.
Note that the stationary domain size decreases as $\mu^{-1}=\delta/\epsilon$
decreases. Obviously, for such a domain to exist its size should be
at least of the order of the front width. This consideration explains
the existence of a critical value $\mu_{st}^{-1}$, below which stationary
domains
do not exist. We note that the transition to stationary patterns beyond
$\mu_{st}^{-1}$ does not occur by a Turing bifurcation. We remind the reader
that we are considering here case $(iii)$ pertaining to bistable systems, and
that none of the homogeneous states of the system lose stability at
the transition point. The transition line in the $\epsilon - \delta$
plane, $\delta=\mu_{st}^{-1}\epsilon$, does resemble, however, the Turing
bifurcation line (2.1).

To study how patterns are affected by the front bifurcation, we begin with
a {\it single} stationary domain in an {\it infinite medium} and follow that
structure as we approach the front bifurcation line. We postpone the discussion
of periodic domain patterns to the next section.
The results to be described below have been obtained
numerically. For more details about the numerical procedures the reader
is referred to section 5. Our starting point in the $\epsilon-\delta$ plane
lies above the transition to stationary domains,
$\delta/\epsilon>\mu_{st}^{-1}$,
and far to the right of the front bifurcation line, namely, at
$\epsilon$ values significantly larger than $\epsilon_c(\delta)$ (see
(3.34) and (3.35)). As $\epsilon$ is decreased (keeping $\delta$ constant)
past a critical value, $\epsilon_{br}(\delta)$, the stationary domain loses
stability in a Hopf bifurcation, and a breathing like oscillatory motion
sets in as shown in Fig. $14a$. This type of oscillations have first been
observed by Koga and Kuramoto [KoK80] and more recently have been studied
analytically by Nishiura and Mimura [NiM89] (see also [KeL90], [OIT90]).
In the latter study a {\it finite} medium has been considered and the
oscillations were found to persist well below $\epsilon _{br}$ without any
indication of a secondary bifurcation.
Our findings for a single domain in an infinite
medium are different; as $\epsilon$ is further decreased the
oscillation amplitude grows, leading eventually to the collapse of the
domain and to a uniform down state as Fig. $14b$ illustrates
(note that the homogeneous states $(u_\pm, v_\pm)$ remain linearly stable
for any $\epsilon$ and $\delta$ values). Similar collapse events have recently
been found in a model for a semiconductor etalon [RRH93].
The collapse can be avoided
by increasing the size of the domain. This can be achieved by making
the system more symmetric, that is, by decreasing $\vert a_0\vert$. For
$\vert a_0\vert$ sufficiently small oscillating domains have been observed
all the way down to the front bifurcation line, $\epsilon=\epsilon_c(\delta)$,
but not below it!

The effect of the front bifurcation on the behavior of a single domain
structure is demonstrated in Figs. $15a$ and $15b$
corresponding to $\epsilon>\epsilon_c(\delta)$ (Ising fronts) and
$\epsilon<\epsilon_c(\delta)$ (Bloch fronts), respectively.
In both figures the initial conditions consist of two fronts bounding a
wide up-state domain and propagating toward one another.
In $15a$ the two fronts set in a stable oscillatory motion, whereas
in $15b$ they rebound from one another and leave
the system in a uniform up state. We attribute this change of behavior to
the appearance of multiple fronts. Above the bifurcation line
($\epsilon>\epsilon_c(\delta))$ there exists only one type of stable front
solution, approaching $(u_\pm,v_\pm)$ as $\chi\to\mp\infty$ and
propagating to the left (assuming $a_0$ negative), or, the symmetric one,
approaching $(u_\mp,v_\mp)$ as
$\chi\to\mp\infty$ and  propagating to the right. Below the bifurcation
($\epsilon<\epsilon_c(\delta))$, another pair of stable front solutions
appear (in addition to the aforementioned one): a front approaching
$(u_\pm,v_\pm)$ as $\chi\to\mp\infty$ and propagating to the right, and  a
front approaching $(u_\mp,v_\mp)$ as $\chi\to\mp\infty$ and propagating to
the left.
In $15b$ the outward propagating front structures converge to the second pair
of stable front solutions and thus proceed with the outward motion
indefinitely. In $15a$, on the other hand, there exist no front solutions
that pertain to an outward motion. The only front solutions that exist
are those propagating toward one another,
and the change in the direction of propagation just reflects
the convergence toward these solutions.

Figs. 16 show another numerical experiment where the same parameter settings
are used but the initial conditions consist now of two fronts propagating
in the same direction. For $\epsilon>\epsilon_c(\delta)$ (Fig. $16a$), the
front on the right side of the domain changes its direction of propagation and
an oscillatory motion sets in. For $\epsilon<\epsilon_c(\delta)$ (Fig. $16b$),
a stable {\it traveling domain}, is obtained.

We note that the stationary domain solution still exists on
both sides of the bifurcation line but is unstable. Fig. 17 shows a typical
bifurcation diagram for domain structures. It does not capture the
oscillatory domain solutions or the collapsing solutions (see section 5)
but it shows the destabilization of the stationary domain solution at
$\epsilon=\epsilon_{br}$, and how traveling domain solutions bifurcate
from it. We also note that by making the system more symmetric
(i.e. decreasing $\vert a_0\vert$) the destabilization of the stationary
domain solution occurs closer to the bifurcation to traveling domains. For
sufficiently symmetric systems
it might be possible for the stationary domain solution to remain stable
all the way down to the traveling-domain branch, but we have not
verified that.

A bifurcation from stationary to traveling domain solutions has already been
found in the context of excitable media [DoK89]. What is new in our findings
is that this transition is a consequence of the front bifurcation. Indeed, we
have verified numerically that the appearance of traveling domain solutions at
$\epsilon=\epsilon_{tr}(\delta)$ (see Fig. 17)
coincides, within the numerical accuracy, with the front
bifurcation line, $\epsilon=\epsilon_c(\delta)$, at least above the onset of
stationary domains (the line $\delta=\mu_{st}^{-1}\epsilon$). Well below that
line $\epsilon_{tr}$ deviates from $\epsilon_c$ and converges, at
$\delta=0$, to $\epsilon_p$ (see section 4.1).

We summarize the discussion of single domain structures in infinite media
by showing in Fig. 18 a phase diagram in the $\epsilon-\delta$ plane.
The diagram is divided into four main regions. Region I,
$\epsilon>\epsilon_{tr}(\delta)$~ and ~$\epsilon>\mu_{st}\delta$, where
domain structures are transient and the asymptotic state is either the uniform
up or uniform down state. Region II, $\epsilon>\epsilon_{br}(\delta)$~ and
{}~$\epsilon<\mu_{st}\delta$, where stationary domains prevails. Region III,
$\epsilon>\epsilon_{tr}(\delta)$,~$\epsilon<\mu_{st}\delta$,~ and
$\epsilon<\epsilon_{br}(\delta)$, where domains oscillate. And region
IV, $\epsilon<\epsilon_{tr}(\delta)$, where domains travel invariably.
The hatched area in region III designates, in a schematic manner, the regime
of steadily oscillating domains. Outside this regime domains oscillate for
a while and then collapse. Depending on the asymmetry of the system, this
regime may shrink to a narrow strip below the line
$\epsilon=\epsilon_{br}(\delta)$ (strong asymmetry), or span the whole
range between the lines $\epsilon=\epsilon_{br}(\delta)$ and
$\epsilon=\epsilon_{tr}(\delta)$ (weak asymmetry). The range between the
two lines, in turn, becomes smaller the more symmetric is the system.
Note the thick solid-dashed line. This is the front bifurcation line
($\epsilon=\epsilon_c(\delta)$) which coincides with the onset of traveling
domains at $\epsilon=\epsilon_{tr}(\delta)$ apart from a small portion at
small $\delta$ values.

\medskip\noindent
{\bf 4.3~ Periodic Domain Patterns}
\smallskip
The behavior of periodic domain patterns is similar in many respects to
that of single domain structures. One notable difference though is
the persistent of stable oscillating domain solutions to the left of the front
bifurcation line ($\epsilon<\epsilon_c(\delta)$), where they coexist with
stable traveling
domain solutions as Figs. 19  demonstrate. The regime of coexistence seems
to be limited; far enough from the front bifurcation
line we found the traveling domain solutions to prevail. The existence of
oscillatory dynamics in the regime of multiple fronts is a consequence
of front interactions. As nearby domains expand and approach one another
there is little room left for diffusion of $v$. As a result the level of
$v$ at any front position increases and the fronts change their directions of
propagation.

Another difference between periodic and single domain patterns is
the possibility of having various modes of oscillation. Ohta {\it et al.}
[OIT90]
have identified three such modes; an in-phase mode where all
domains expand and shrink simultaneously, an out-of-phase mode where nearby
domains oscillate with a phase shift of $\pi$, and an ``acoustic''
mode, where nearby domains travel in opposite directions
back and forth (note that in the latter case the spaces between the domains,
or the down-state domains, undergo out-of-phase oscillations).
Of the three, the in-phase mode is the first to be excited. As pointed out
by Ohta {\it et al.} [OIT90], this can be understood by considering the
stabilizing effect the diffusion of $v$ has. The in-phase mode involves
uniform deviations of $v$ from the stationary domain profile
and stabilization by diffusion does not take place.
The out-of-phase mode, on the other hand, pertains to alternating deviations
every second front site. In this case diffusion of $v$ tends to smear out
these deviations and thus to stabilize the stationary pattern against this
mode. Similar considerations hold for the acoustic mode. In the numerical
experiments we have conducted the in-phase mode always prevailed, even at
relatively small $\delta$ values, where the diffusive damping of that mode
is weaker.

We expect Hopf-Turing systems, driven far beyond the Turing instability,
to exhibit the same phenomenology that short-wavelength patterns in
bistable systems do. Indeed, by decreasing $\epsilon$ at constant $\delta$
(far beyond the Turing instability) we found a transition from stationary
to oscillating patterns, followed by a transition to traveling patterns.

\medskip\noindent
{\bf 4.4~ Reflection at Boundaries}
\smallskip
The front bifurcation may drastically affect the behavior of fronts near
an impermeable boundary. The effect of such a boundary is similar
to that of another approaching front and the scenario
of behaviors is similar to that found for domain structures.
Figs. 20  show the boundary effects as we cross
all four regions in the $\epsilon-\delta$ plane by decreasing $\epsilon$ at
constant $\delta$ (the simulations were carried out with Neumann boundary
conditions). In region I (transient domains) the front is absorbed
at the boundary, leaving the system in a uniform state. In region II
(stationary domains), the front comes to a stop at a characteristic
distance from the boundary, whereas in region III (oscillating domains) it
oscillates near the boundary. In all these regions the (Ising) front is either
absorbed at the boundary or is bound to it. By contrast, in region IV
the front is {\it reflected}. This is a consequence of the coexistence of two
counter propagating Bloch fronts beyond the front
bifurcation line. At front speeds high enough relative
to the diffusion of $v$, both Ising and Bloch fronts are absorbed at the
boundaries. Ising fronts, however, never reflect at boundaries.

\medskip\noindent
{\bf 5.~ Numerical Procedures}
\smallskip
Numerical simulations of (1.1) were performed using the method of lines with
the spatial derivatives approximated by 2nd or 4th order finite differences on
a uniform mesh.  To solve the resulting system of ordinary differential
equations, we used a stiff ODE solver [NKM89]. This solver implements the
implicit methods of Gear [Gea71] with adaptive control of both time step
and method order.

For the computations pertaining to single domain structures in inifinite
systems the simulations were
done with Neumann (no flux) boundary conditions and large computational system
sizes.  We checked the results with larger system sizes to ensure that the
boundaries did not contribute any measurable effect on the solution.  When the
simulations pertain to periodic patterns, periodic
boundary conditions were applied.  Occasionally we used periodic boundary
conditions for single traveling domains rather than Neumann boundary
conditions (see for example Fig. 16$b$). In those cases we chose the period
to be large in comparison with the domain size and checked that the
two types of boundary conditions gave the same solution within the numerical
accuracy.

For the continuation of solutions we used the bifurcation package AUTO
[Doe86].  Starting with a stable traveling or stationary solution computed
with our PDE solver we continued a periodic solution in one or two parameters
to study the effects of parameter variations on the existence and stablilty of
solutions.  When AUTO failed to give accurate stability information, due to the
approach of the periodic orbit towards two heteroclinic cycles, we used
our PDE solver to integrate the computed solution forward in time to check its
stability.

To find the oscillating domains region we linearized (1.1) about the
numerically computed stationary solution and solved the resulting eigenvalue
problem.  The spatial part of the linearized operator was discretized using 2nd
order finite differences on a nonuniform mesh with grid points concentrated
near the steep gradient front structures of the domain.  We solved the
eigenvalue problem using the IMSL routine DEVCRG for real general matrices.
Decreasing $\epsilon$ below the line $\epsilon=\epsilon_{br}(\delta)$ a pair of
complex conjugate eigenvalues crosses the imaginary axis indicating that the
stationary stable solution goes unstable through a Hopf bifurcation to an
oscillating domain.

\medskip\noindent
{\bf 6.~ Conclusion}
\smallskip
We have studied a front bifurcation, likely to occur in a variety of
nonvariational reaction diffusion systems, and the possible consequences
of this bifurcation on pattern formation and pattern dynamics. The
bifurcation is accompanied
by the appearance of a second independent field which breaks the odd symmetry
of front solutions, and leads to a richer dynamical behavior.
The main effect of the bifurcation, regarding patterns, is to
allow the formation of stable
traveling domain patterns. For small $\delta$ values (including $\delta=0$),
it provides a mechanism for pattern formation as no persistent domain
patterns can be formed below the bifurcation. For larger $\delta$ values
it is responsible for transitions from transient, oscillating, or stationary
patterns to traveling ones. Other consequences of the bifurcation are a
coexistence regime of traveling and oscillating patterns, and the possible
reflection of fronts at boundaries.

These results were obtained for bistable systems but are also relevant to
excitable and Hopf-Turing systems as the latter can be obtained by
unfolding the former. In this sense, the variety of chemical-wave
phenomena in excitable and oscillatory systems can be regarded as a
consequence of an Ising-Bloch front bifurcation occurring nearby in
parameter space. A more specific implication for Hopf-Turing systems is the
prediction of a transition from
stationary to oscillating patterns, and from oscillating to traveling
patterns  far beyond the Turing instability. These transitions may shed light
on the complex spatio-temporal behavior that has recently been observed
in the CIMA reaction by Ouyang and Swinney [OuS91].

Beyond these specific results and implications, the present analysis
demonstrates the usefulness of focusing attention on localized structures
that appear in extended patterns. This view lies behind the various kinematical
approaches to pattern dynamics [KiM89,NeM92,Mer92,CrH93], where the effects
of small perturbations and weak interactions on the dynamics of localized
structures are studied. The point we wish
to emphasize here is that it is also
important to look for structural changes that localized structures might
undergo as these can drastically affect pattern dynamics.

We confined ourselves in
this paper to patterns in one space dimension, but we expect the front
bifurcation to have important effects in two dimensions as well.
Perhaps the first question that needs to be addressed concerns the
dynamics of a curved Bloch front near the front bifurcation point.
The dynamics of a such a front far beyond the bifurcation is well known
[TyK88,Mer92]; the normal speed $c_n$ is affected by the curvature $k$
according to the simple linear law, $c_n=c-Dk$, where $D$ is a positive
constant having the dimension of a diffusion coefficient. The consequent
effect of curvature is to smooth out any wiggles along the front and thus to
stabilize its dynamics.
Close to the bifurcation point we expect this relation to be highly
nonlinear. This might change the role of curvature in a way that still needs
to be explored.

An important consequence of the Ising-Bloch front bifurcation in two
space dimensions is the possible formation of spiral waves. As pointed
out by Coullet, a spiral wave can be created by joining end to end two
counter propagating Bloch fronts. This view goes back
to Fife [Fif88] who discussed the twisting action induced by
joining a front and a back end to end in an excitable medium.
All studies of spiral waves so far have taken $\epsilon$ to be a small
parameter, and therefore are only valid far from the front bifurcation. The
expected nonlinear relation between curvature and normal speed close to the
bifurcation point makes the study of spiral waves in that regime
potentially interesting.

%We conclude with a remark on the extended FHN model (1.1). Not only is
%this model generic for reaction diffusion systems, such as chemical
%reactions, excitable membranes, etc., it is also closely related
%to pattern formation models for near equilibrium systems characterized
%by a nonconserved order parameter.

\vskip .4 truein
\centerline{\bf Acknowledgements}
\medskip
We would like to thank Rob Indik and Jerry Moloney for interesting
discussions. We also
wish to thank the Arizona Center for Mathematical Sciences (ACMS)
for support. ACMS is sponsored by AFOSR contract FQ8671-9000589
(AFOSR-90-0021) with the University Research Initiative Program at the
University of Arizona. One of us (A.H.) acknowledges the support of
the Computational Science Graduate Fellowship Program of the Office of
Scientific Computing in the Department of Energy.

\vfil
\eject
\noindent
{\bf Appendix A: The front bifurcation for $\epsilon/\delta\ll 1$}
\medskip
We consider here the regime $\epsilon/\delta\ll 1$ and use singular
perturbation theory to derive the front bifurcation line. Rescaling
space and time according to
%******************************* A1 ******************************
$$z=\sqrt\mu x,~~~~~~~~\tau=\epsilon t,~~~~~~~~\mu=\epsilon/\delta\ll 1,$$
and assuming traveling solutions, $u=u(\zeta)$ and $v=v(\zeta)$, where
$\zeta=z-c\tau$, we obtain from (1.1)
%******************************* A2 ********************************
$$\eqalign
{\mu u_{\zeta\zeta}+c\delta\mu u_\zeta+u-u^3-v =0\cr
v_{\zeta\zeta}+cv_\zeta+u-a_1v-a_0=0.\cr}\eqno(A2)$$
We consider front solutions of $(A2)$ that satisfy  $(u,v)\to(u_\pm,v_\pm)$
as $\zeta\to \mp\infty$, and set the origin, $\zeta=0$, by
demanding $u(0)=0$. Outer solutions of $(A2)$ are obtained by setting
$\mu=0$ and solving
%******************************** A3 *********************************
$$v_{\zeta\zeta}+cv_\zeta+u_\pm(v)-a_1v-a_0=0,\eqno(A3)$$
for $\zeta<0$ ($u=u_+(v)$) and $\zeta>0$ ($u=u_-(v)$).
To simplify $(A3)$ we approximate the branches $u_\pm(v)$ by the
linear forms (4.3).
This approximation can be controlled by varying $a_1$. We then obtain
the following boundary value problems:
%******************************** A4 **********************************
$$\eqalign{v_{\zeta\zeta}+cv_\zeta-k^2v+k^2v_+=0,~~~~~~~~\zeta<0\cr
v(-\infty)=v_+~~~~~~~~~~v(0)=v_f,\cr}\eqno(A4)$$
and
$$\eqalign{v_{\zeta\zeta}+cv_\zeta-k^2v+k^2v_-=0,~~~~~~~~\zeta>0\cr
v(\infty)=v_-~~~~~~~~~~v(0)=v_f,\cr}\eqno(A5)$$
where $v_\pm$ are given by (4.4$b$),~ $k^2=a_1+{1\over 2}$ and $v_f$
is the level of $v$ at the front position. The solutions are
%********************************* A6,7 *********************************
$$v(\zeta)=(v_f-v_+)e^{\sigma_1\zeta}+v_+,~~~~~~~~\zeta<0,\eqno(A6)$$
$$v(\zeta)=(v_f-v_-)e^{\sigma_2\zeta}+v_-,~~~~~~~~\zeta>0,\eqno(A7)$$
where
%********************************* A8 ************************************
$$\sigma_{1,2}=-c/2\pm(c^2/4+k^2)^{1/2}.\eqno(A8)$$
Matching the derivatives of the two outer solutions at $\zeta=0$ yields
%********************************* A9 ************************************
$$v_f=-{c\over 2k^2(c^2/4+k^2)^{1/2}}-{a_0\over k^2}.\eqno(A9)$$
A second relation between $v_f$ and $c$ is obtained by solving the inner
problem. To this end we stretch the traveling-frame coordinate according
to $\chi=\zeta/\sqrt\mu$ and obtain from ($A2$)
%********************************** A10 **********************************
$$\eqalign{
u_{\chi\chi}+\eta c u_\chi+u-u^3-v=0,\cr
v_{\chi\chi}+\sqrt\mu cv_\chi+\mu(u-a_1 v-a_0)=0,\cr}\eqno(A10)$$
where $\eta^2=\epsilon\delta$.
Setting formally $\mu=0$ in ($A10$) we obtain a nonlinear
eigenvalue problem for $c$,
%********************************** A11 **********************************
$$\eqalign{
u_{\chi\chi}+\eta c u_\chi+u-u^3-v_f=0,\cr
u(\mp\infty)=u_\pm(v_f)=\pm 1-v_f/2,\cr}\eqno(A11)$$
which yields the second relation between $v_f$ and $c$:
%********************************** A12 **********************************
$$v_f=-{\sqrt 2\over 3}\eta c.\eqno(A12)$$
Comparing ($A9$) and $(A12)$  we find
%********************************** A13 **********************************
$${\sqrt 2\over 3}\eta c= {c\over 2k^2(c^2/4+k^2)^{1/2}}+{a_0\over k^2}.
\eqno(A13)$$
A plot of the solutions  $c=c(\eta)$ of ($A13$) in the $(c,\eta)$ plane
yields a bifurcation diagram with $\eta$ being the bifurcation parameter.
Consider now the symmetric case, $a_0=0$. Assuming propagating solutions
($c\ne 0$) and taking the limit $c\to 0$ we find the bifurcation point
%*******************************************************************
$$\eta_c={3\over(2a_1+1)^{3/2}},$$
or, recalling that $\eta^2=\epsilon\delta$, the bifurcation line
%********************************** A14 ******************************
$$\delta_c(\epsilon)={9\over \epsilon(2a_1+1)^3}.\eqno(A14)$$

\vfil
\eject
\noindent
{\bf Appendix B: Stationary-Domain Solutions for $\epsilon/\delta\ll 1$}
\medskip
Stationary-domain solutions of (1) satisfy
%************************* B1 **************************************
$$\eqalign
{\mu u_{zz}+u-u^3-v =0\cr
v_{zz}+u-a_1v-a_0=0,\cr}\eqno(B2)$$
where $z=\sqrt\mu x$ and $\mu=\epsilon/\delta$ is sufficiently small.
In the following we will assume that $\mu\ll 1$ and that $a_0<0$. The
latter assumption means that an Ising front connecting an up state
on the left with a down state on the right propagates to the left
and consequently that a single domain is an up-state domain (see Fig. 13).
Outer solutions are obtained by setting $\mu=0$ in ($B2$). We look for
such solutions in three regions, $z<z_l$, $z>z_r$, and $z_l<z<z_r$, where
$z_l$ and $z_r$ are the positions of the fronts that bound the stationary
domain from the left and the right, respectively. The first two regions
pertain to down-state domains in which we need to solve
%************************** B2 **************************************
$$v_{zz}-k^2v+k^2v_-=0,\eqno(B2)$$
subject to the boundary conditions
%*************************** B3 ***************************************
$$\eqalign{
v(-\infty)=v_- ~~~~~~~~~~ v(z_l)=v_l
{}~~~~~~~~~~{\rm for}~~~~~z<z_l,\cr
v(\infty)=v_- ~~~~~~~~~~  z(z_r)=v_r
{}~~~~~~~~~~{\rm for}~~~~~z>z_r,\cr}\eqno(B3)$$
where $v_l$ and $v_r$ are yet to be determined.
The third region, $z_l<z<z_r$, corresponds to an up-state domain in which
we solve
%**************************** B4 ****************************************
$$v_{zz}-k^2v+k^2v_+=0,\eqno(B4)$$
subject to the boundary conditions
%****************************  B5 **************************************
$$v(z_l)=v_l ~~~~~~~~~~ v(z_r)=v_r.\eqno(B5)$$
In $(B2)$ and $(B4)$, $k^2=a_1+1/2$
and we have used the linearized forms of the branches $u_\pm(v)$ (see
(4.3)).

The solutions of $(B2)$ to $(B5)$ are:
%*******************************  B6,7 ***********************************
$$v(z)=(v_l-v_-)e^{k(z-z_l)}+v_- ~~~~~~~~~~~~~~~~~~~~~~~~~~z<z_l,\eqno(B6)$$
$$v(z)=(v_r-v_-)e^{-k(z-z_r)}+v_- ~~~~~~~~~~~~~~~~~~~~~~~~z>z_r,\eqno(B7)$$
and
%******************************  B8  ***********************************
$$\eqalign{v(z)={1\over \sinh(k\Lambda)}\Bigl[(v_r-v_+)\sinh k(z-z_l)-
(v_l-v_+)\sinh k(z-z_r)\Bigr]+v_+\cr
{}~~~~~~~~~~~~~~~~~~~~~~~~~~~~~~~~~~~~~~~~~~~~~~~~~~~~~~~~~~~~~~~~~~~~~~~~~~~
z_l<z<z_r,\cr}\eqno(B8)$$
where $\Lambda=z_r-z_l$.
Matching  the first derivatives of these solutions  at $z=z_l$
and $z=z_r$ we find that $v_l=v_r$ and that $\Lambda$ (the domain width)
solves the equation
%******************************  B9 ***********************************
$$\coth k\Lambda-{\rm csch} k\Lambda=(v_r-v_-)/(v_+-v_r).\eqno(B9)$$

Equation $(B9)$ still contains one unknown parameter, $v_r$. This parameter
is obtained from the analysis of the inner regions centered at $z=z_l$
and $z=z_r$. Stretching the coordinate system according to $x=z/\sqrt\mu$
and setting $\mu=0$ we find
%*******************************  B10  ***************************
$$u_{xx}+u-u^3-v_r=0,\eqno(B10)$$
implying $v_r=0$. Using this result in ($B9$) and solving for $\Lambda$
we get
%******************************* B11  ********************************
$$\Lambda=-{1\over k}\ln(-a_0),\eqno(B11)$$
or (4.13) where $\lambda=\Lambda/\sqrt\mu$.

\vfil\eject
\centerline{\bf References}
\medskip
\item{[CaP89]} J. Carr and R. L. Pego 1989 {\it Comm. Pure and Appl. Math.}
{\bf 42} 523.
\item{[CDB90]} V. Castets, E. Dulos, J. Boissonade and P. De Kepper 1990
{\it Phys. Rev. Lett.} {\bf 64} 2953.
\item{[CLH90]} P. Coullet, J. Lega, B. Houchmanzadeh, and J. Lajzerowicz,
1990 {\it Phys. Rev. Lett.} {\bf 65} 1352.
\item{[CrH93]} M. C. Cross and P. C. Hohenberg 1993 Pattern Formation
Outside of Equilibrium, preprint.
\item{[DKT88]} J. D. Dockery, J. P. Keener and J. J. Tyson 1988
{\it Physica D} {\bf 30} 177.
\item{[Doe86]} E. J. Doedel 1986 {\it Software for continuation and
bifurcation problems in  ordinary differential equations} California
Instititue of Technology.
\item{[DoK89]} J. D. Dockery and J. P. Keener 1989 {\it SIAM J. Appl. Math.}
{\bf 49} 539.
\item{[EHT84]} G. B. Ermentrout, S. P. Hastings and W. C. Troy 1984 {\it
SIAM J. Appl. Math.} {\bf 42} 219.
\item{[EMS88]} C. Elphick, E. Meron and E. A. Spiegel 1988 {\it Phys. Rev.
Lett.} {\bf 61} 496.
\item{[FiB85]}  R. J. Field and M. Burger 1985 {\it Oscillations and Traveling
Waves in Chemical Systems} Wiley, New York.
\item{[Fif85]} P. C. Fife 1985 {\it J. Stat. Phys.} {\bf 39} 687.
\item{[Fif88]} P. C. Fife 1988 {\it CBMS-NSF Regional Conf. Series in
Appl. Math.} {\bf 53} 1.
\item{[Gea71]} C. W. Gear 1971 {\it Numerical initial value problems in
ordinary differential equations} Prentice-Hall.
\item{[IMN89]} H. Ikeda, M. Mimura and Y. Nishiura 1989 {\it Nonl. Anal.
TMA} {\bf 13} 507.
\item{[KaO82]} K. Kawasaki and T. Ohta 1982 {\it Physica} {\bf 116A} 573.
\item{[Kee80]} J. P. Keener 1980 {\it SIAM J. Appl. Math.} {\bf 39} 528.
\item{[KoK80]} S. Koga and Y. Kuramoto 1980 {\it Prog. Theor. Phys.} {\bf 63}
106.
\item{[KeL89]} D. A. Kessler and H. Levine 1989 {\it Physica D} {\bf 39} 1.
\item{[KeL90]} D. A. Kessler and H. Levine 1990 {\it Phys. Rev. A} {\bf 41}
5418.
\item{[KiM89]} Y. S. Kivshar and B. A. Malomed 1989 {\it Rev. Mod. Phys.}
{\bf 61} 763.
\item{[LaN79]} J. Lajzerowicz and J. J. Niez 1979 {\it J. Phys (Paris) Lett.}
{\bf 40} L165.
\item{[LeE92]} I. Lengyel and I. R. Epstein 1992 {\it Proc. Natl. Acad.
Sci. U.S.A.} {\bf 89} 3977.
\item{[LKE92]} I. Lengyel, S. K\'ad\`ar, and I. R. Epstein 1992 {\it Phys.
Rev. Lett.} {\bf 69} 2729.
\item{[Mer92]} E. Meron 1992 {\it Physics Reports} {\bf 218} 1.
\item{[Mur89]} J. D. Murray 1989 {\it Mathematical Biology} Springer, New
York.
\item{[NeM92]} A. C. Newell and J. V. Moloney 1992 {\it Nonlinear Optics}
Addison Wesley (section 3g).
\item{[NiM89]} Y. Nishiura and M. Mimura 1989 {\it SIAM J. Appl. Math.}
{\bf 49} 481.
\item{[NKM89]} S. Nash, D. Kahaner and C. Moler 1989 {\it Numerical methods
and software} Prentice Hall.
\item{[OIT90]} T. Ohta, A. Ito and A. Tetsuka 1990 {\it Phys. Rev. A} {\bf 42}
3225.
\item{[OrR75]} P. Ortoleva and J. Ross 1975 {\it J. Chem. Phys.} {\bf 63}
3398.
\item{[OuS91]} Q. Ouyang, H. L. Swinney 1991 {\it Chaos} {\bf 1} 411.
\item{[RiT82]} J. Rinzel and D. Terman 1982 {\it SIAM J. Appl. Math.}
{\bf 42} 1111.
\item{[RoM92]} A. Rovinsky and M. Menzinger 1992 {\it Phys. Rev. A} {\bf 46}
6315.
\item{[RRH93]} U. A. Rzhanov, H. Richardson, A. A. Hagberg, and J. V. Moloney
1993 {\it Phys. Rev. A} {\bf 47} 1480.
%\item{[SkS91]} G. S. Skinner and H. L. Swinney {\it Physica D} {\bf 48} 1.
\item{[Tur52]} A. M. Turing 1952 {\it Philos. Trans. Roy. Soc. London B}
{\bf 327} 37.
\item{[TyK88]} J. J. Tyson and J. P. Keener 1988 {\it Physica D} {\bf 32}
327.
%\item{[Win72]} A. T. Winfree 1972 {\it Science} {\bf 175} 634.
%\item{[ZaZ70]} A. N. Zaikin and A. M. Zhabotinsky 1970 {\it Nature} {\bf 225}
%535.

\vfil\eject

\noindent
\centerline{\bf Figure Captions}
\medskip\noindent
{\bf Figure 1:}~
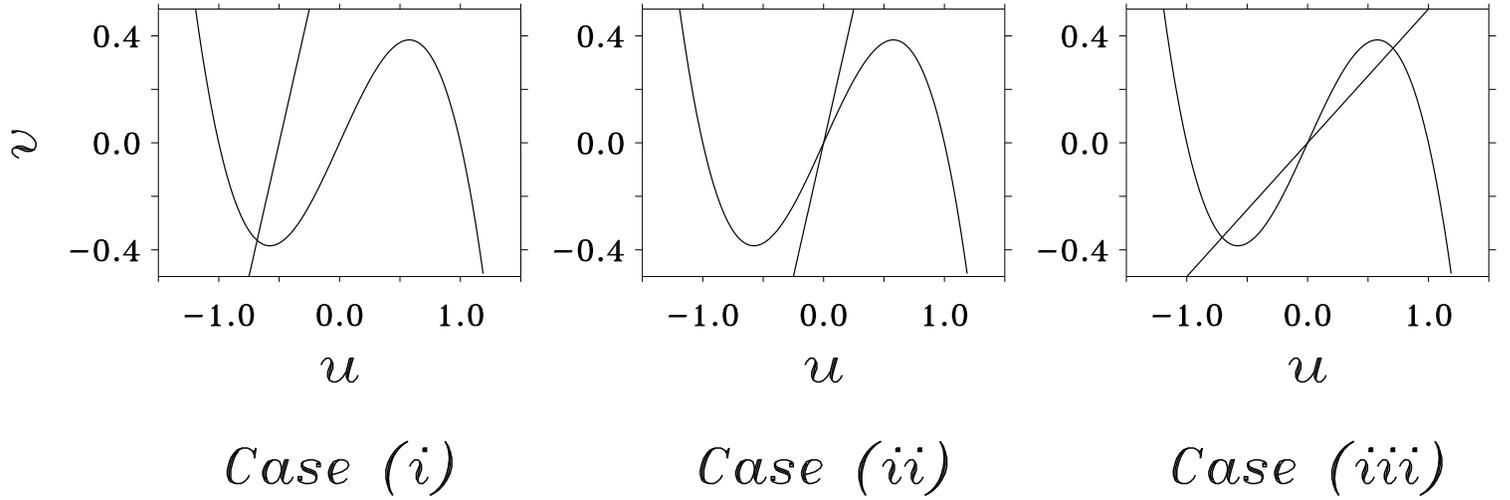
Three basic cases for the nullclines of the model system (1.1). Case
(i): The nullclines intersect at a single point on an outer branch of the cubic
nullcline. Case (ii): The nullclines intersect at a single point on the middle
branch of the cubic nullcline.  Case (iii): The nullclines intersect at three
points each on a different branch of the cubic nullcline.\hfill
\break
{\bf Figure 2:}~
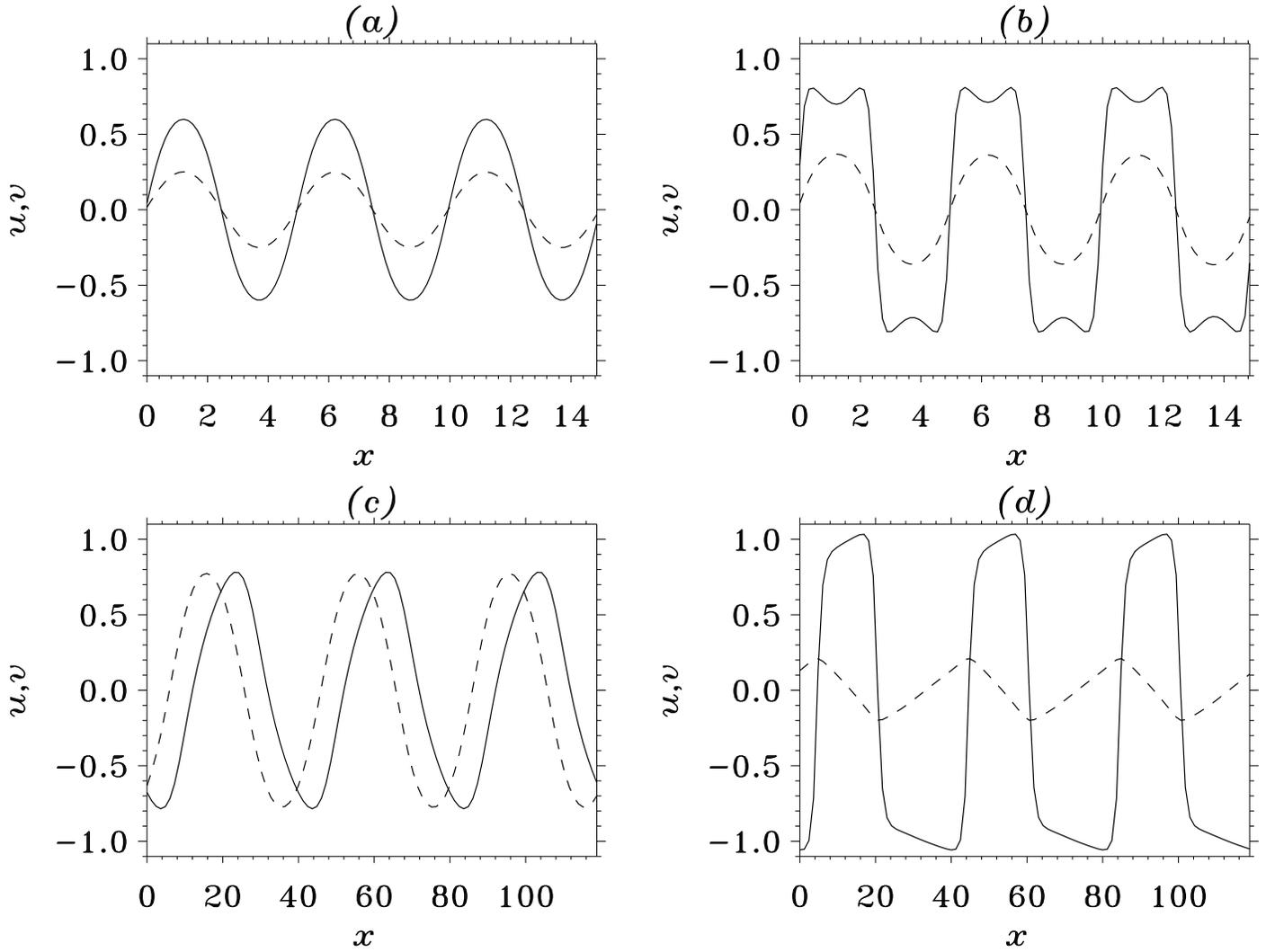
Patterns near and far from onset show the development
of two spatial scales. (a) Stationary periodic pattern
near onset of the Turing instability.  (b) Stationary periodic pattern far
from onset of the Turing instability.  (c) Traveling pattern near the Hopf
bifurcation.  (d) Traveling pattern far from the Hopf bifurcation.  In
all the figures the solid line is the $u$ field and the dashed line is
the $v$ field.\hfill
\break
{\bf Figure 3:}~
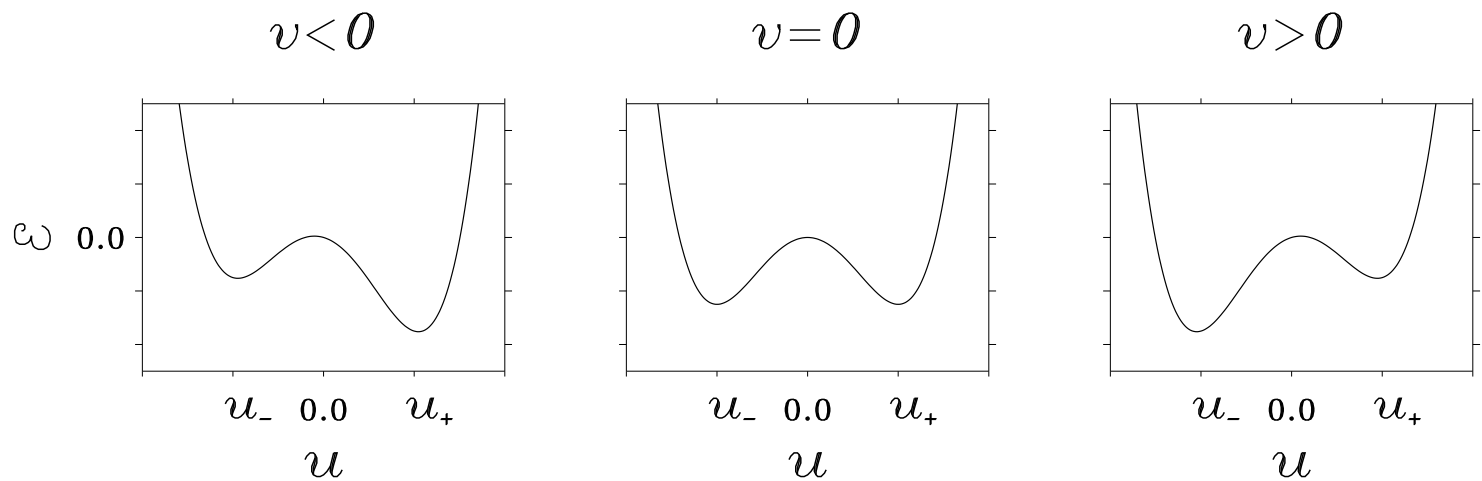
The free energy density ${\cal E}$ (see (3.3)) for different values of
$v$.\hfill
\break
{\bf Figure 4:}~
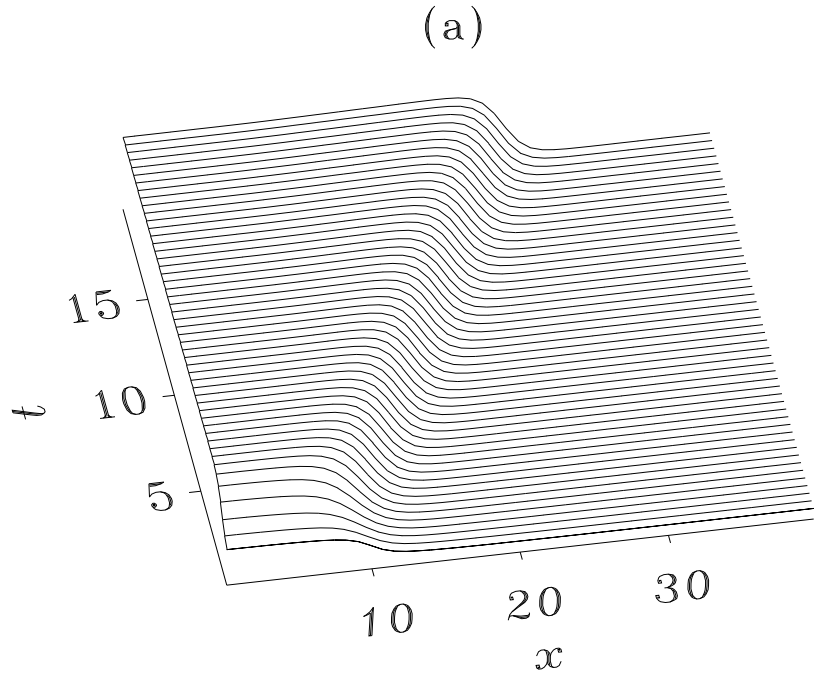
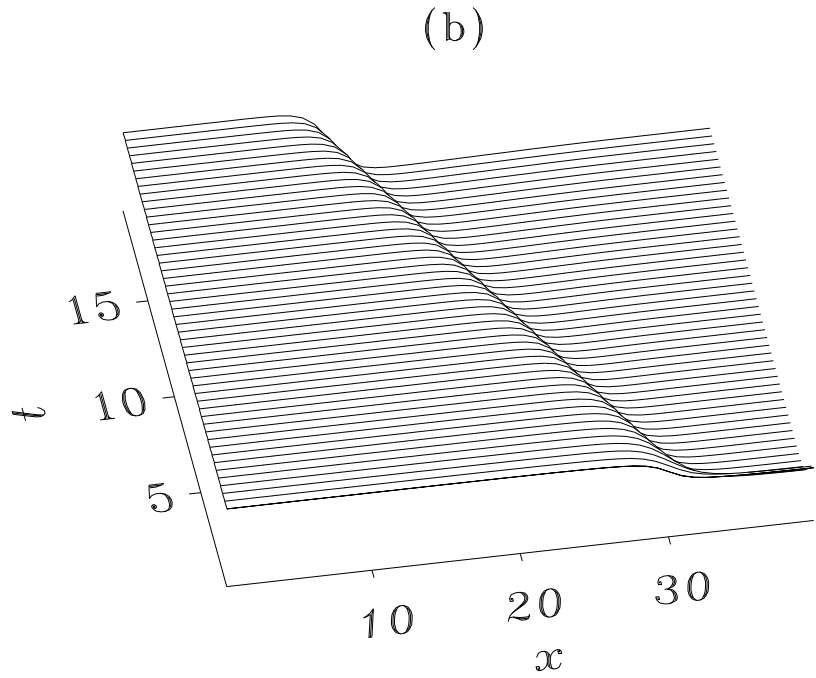
Multiplicity of stable front solutions
when $\epsilon \ll 1$. (a) Preparing the system in the down state and
perturbing it locally at the left edge of the domain induces a front
propagating to the right. (b) Preparing the system in the up state and
perturbing it locally at the right edge of the domain induces a front
propagating to the left.\hfill
\break
{\bf Figure 5:}~
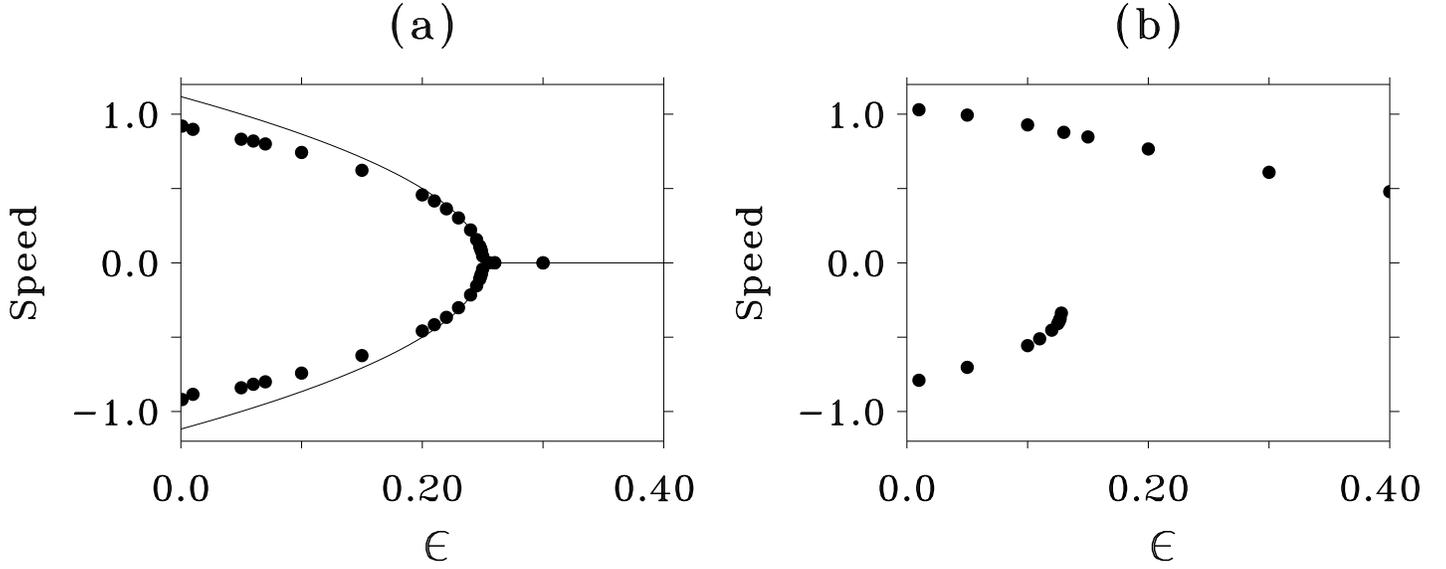
Bifurcation diagrams of front solutions. The dots are data
points representing  the speed of the different types of stable front solutions
that exist for each value of $\epsilon$.
(a) The symmetric case ($a_0=0$). The solid line is the theoretical
bifurcation diagram computed from (3.21).
(b) The nonsymmetric case ($a_0=.1$) shows the unfolding of the pitchfork to
a saddle node bifurcation.\hfill
\break
{\bf Figure 6:}~
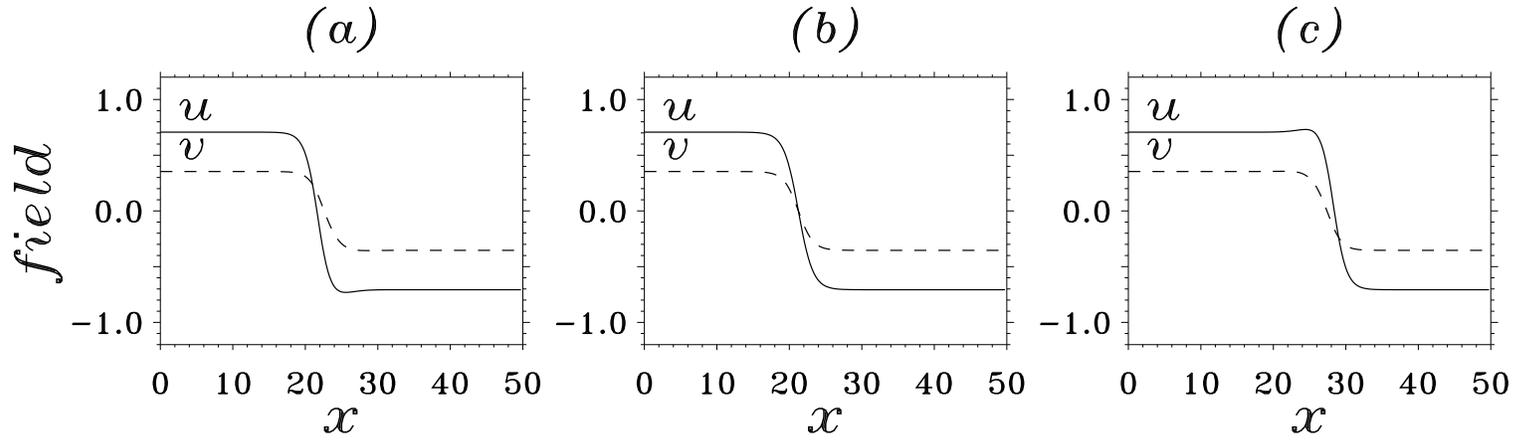
The propogation direction of Bloch fronts is determined
by the translation of the $v$ field relative to the $u$ field; $v$ always
lags behind $u$ unless the front is stationary.
(a) Left traveling front.
(b) Stationary front. (c) Right traveling front.\hfill
\break
{\bf Figure 7:}~
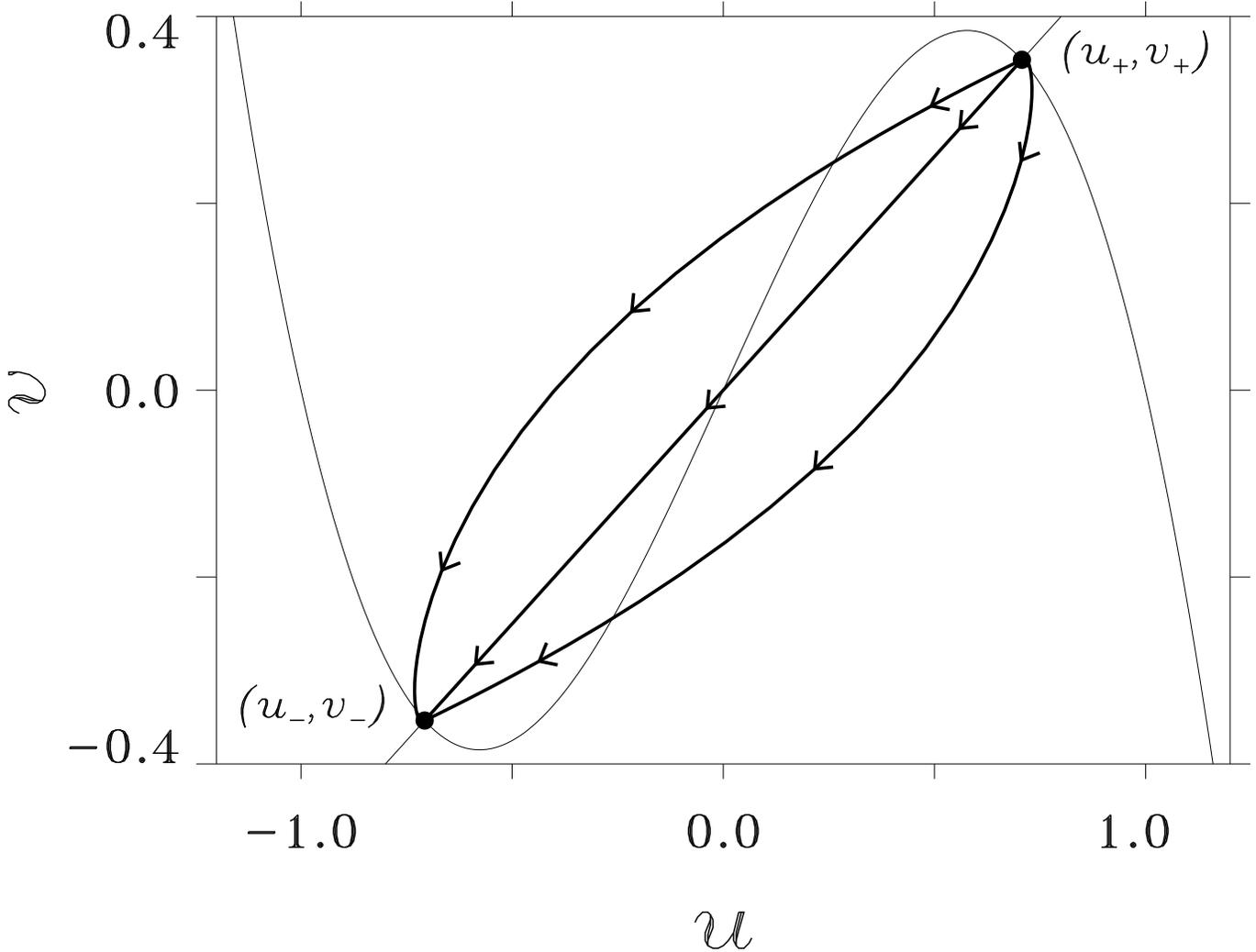
Phase portraits of front solutions connecting the $(u_+,v_+)$
state at $\chi=-\infty$ to the $(u_-,v_-)$ state at $\chi=\infty$ for the
symmetric model ($a_0=0$) with $\delta=0$.
The light colored curves are the nullclines $f=0$ and $g=0$ and the dark
colored curves are the numerically computed trajectories.
The trajectory through $(0,0)$ corresponds to the symmetric stationary
front solution.
The two other trajectories break the odd symmetry of the system and
correspond to traveling fronts.
The computational parameters are $\epsilon=1.0,\delta=0,a_1=2.0,a_0=0$ for
the stationary front and the same with $\epsilon=.2$ for the two traveling
fronts.\hfill
\break
{\bf Figure 8:}~
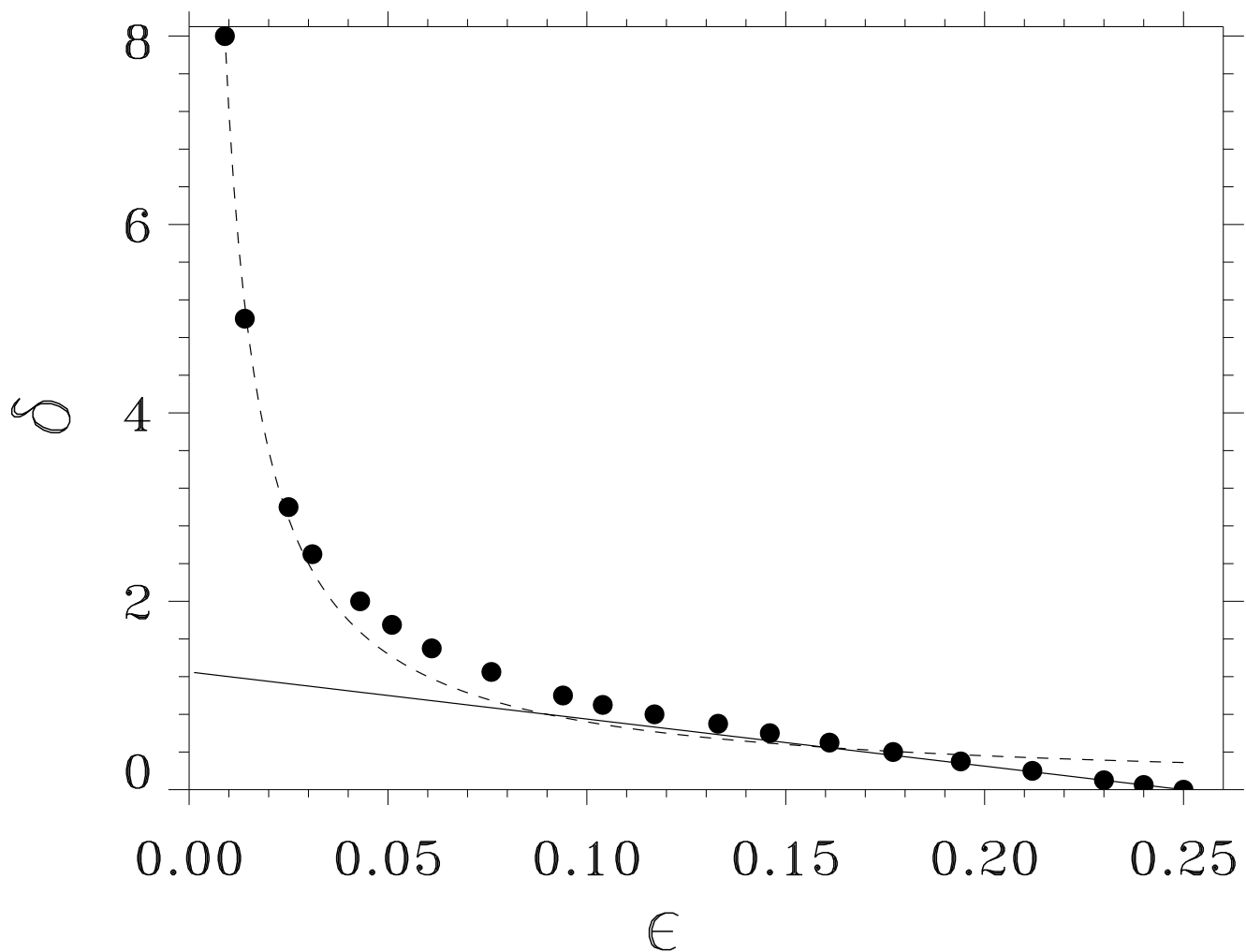
The front bifurcation line in the $\epsilon - \delta$ plane for
the symmetric ($a_0=0$) model with $a_1=2.0$.  The solid circles represent
the numerically calculated bifurcation line.  The solid line
is the linear approximation (valid for $\delta \ll 1$) and the dashed
line is the singular pertubation result (valid for $\delta / \epsilon \gg
1$).\hfill
\break
{\bf Figure 9:}~
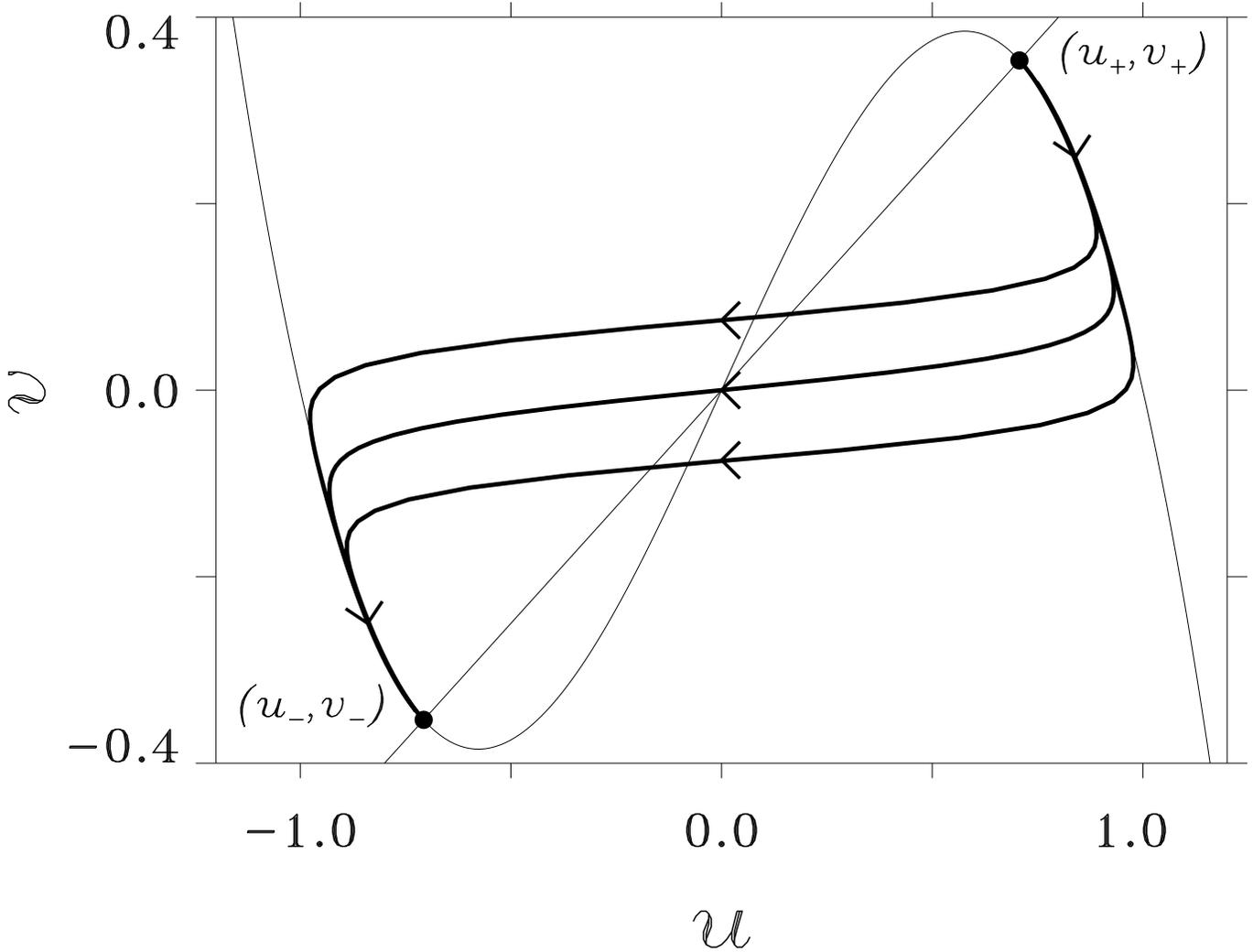
 Phase portraits of front solutions connecting the
$(u_+,v_+)$ state at $\chi=-\infty$ to the $(u_-,v_-)$ state at
$\chi=\infty$ for the symmetric model ($a_0=0$) with $\delta>1$.
The light colored curves are the nullclines $f=0$ and $g=0$ and the dark
colored curves are the numerically computed trajectories.
The trajectory through $(0,0)$ corresponds
to the symmetric stationary front with $\epsilon > \epsilon_c(\delta)$ and
the two other trajectories correspond to the symmetry breaking traveling
fronts with $\epsilon <\epsilon_c(\delta)$. \hfill
\break
{\bf Figure 10:}~
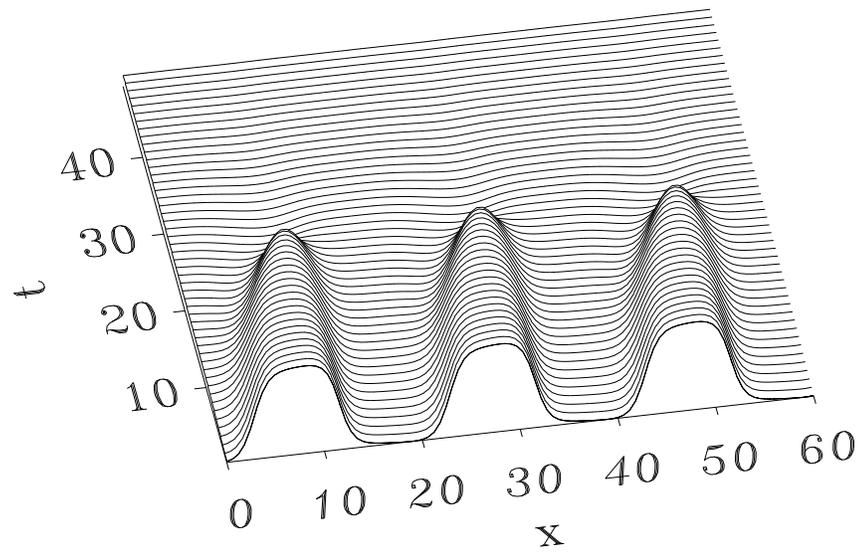
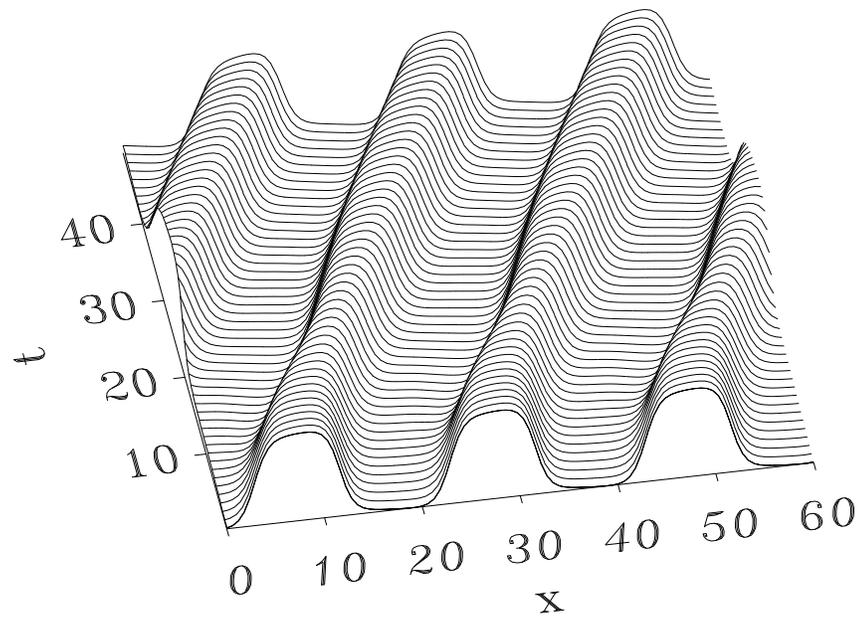
(a) Relaxation of an initial pattern to the uniform down state for
the regime of Ising fronts, $\epsilon=0.14, \delta=0, a_1=2.0, a_0=-0.1$.
(b) The convergence of the same initial pattern toward a stable traveling
pattern beyond the bifurcation from Ising
to Bloch fronts, $\epsilon=0.09$.\hfill
\break
{\bf Figure 11:}~
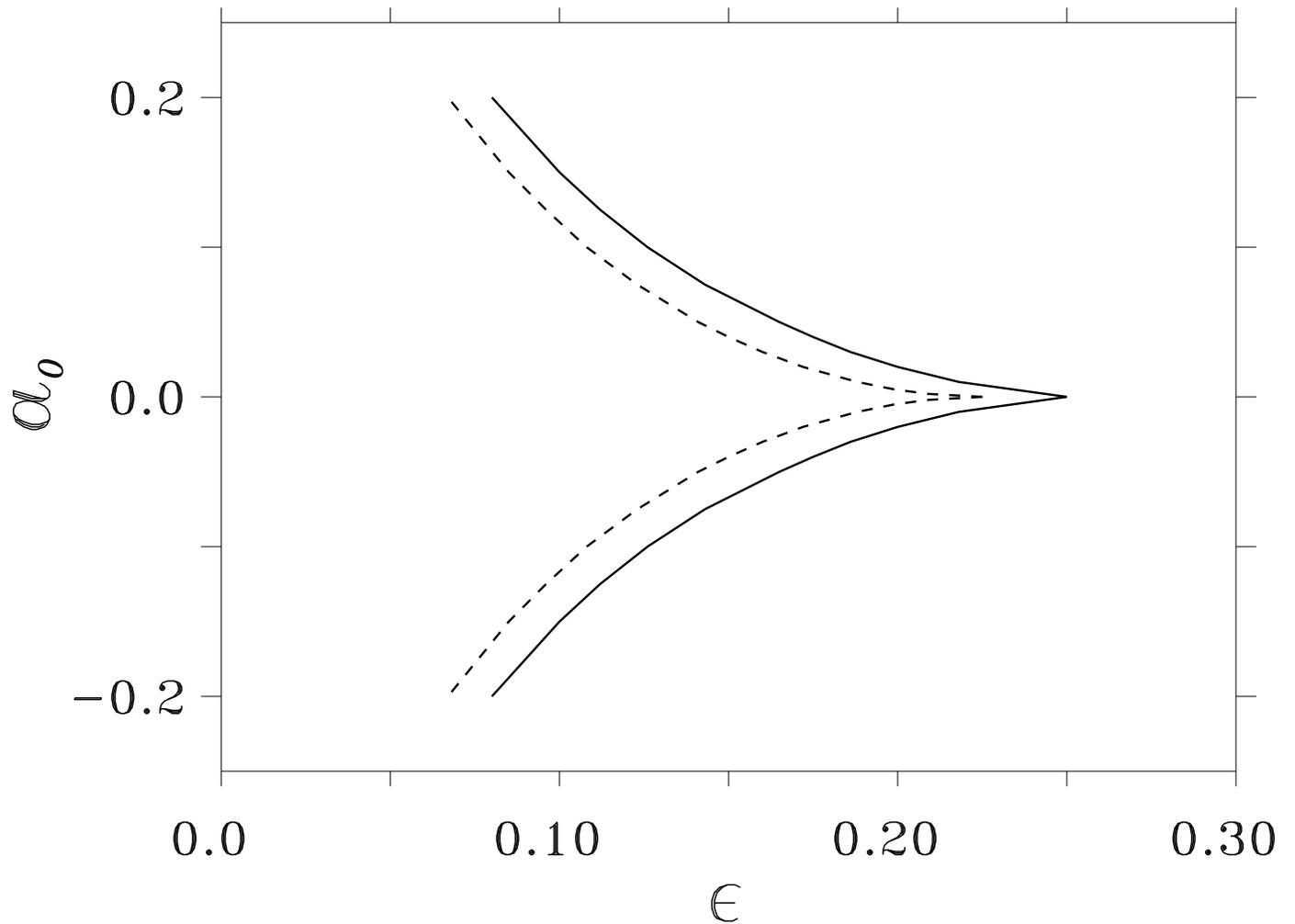
 Phase diagram in the $\epsilon-a_0$ plane. For the region
to the right of the solid curve only one type of front solution exists and
initial patterns do not persist. In the region between the solid and the dashed
curves multiple stable fronts coexist but patterns still decay toward a uniform
state. For the region to the left of the dashed curve initial patterns evolve
toward persistent patterns in the form of stable traveling waves. Computational
parameters are : $\delta=0,~ a_1=2.0$.\hfill
\break
{\bf Figure 12:}~
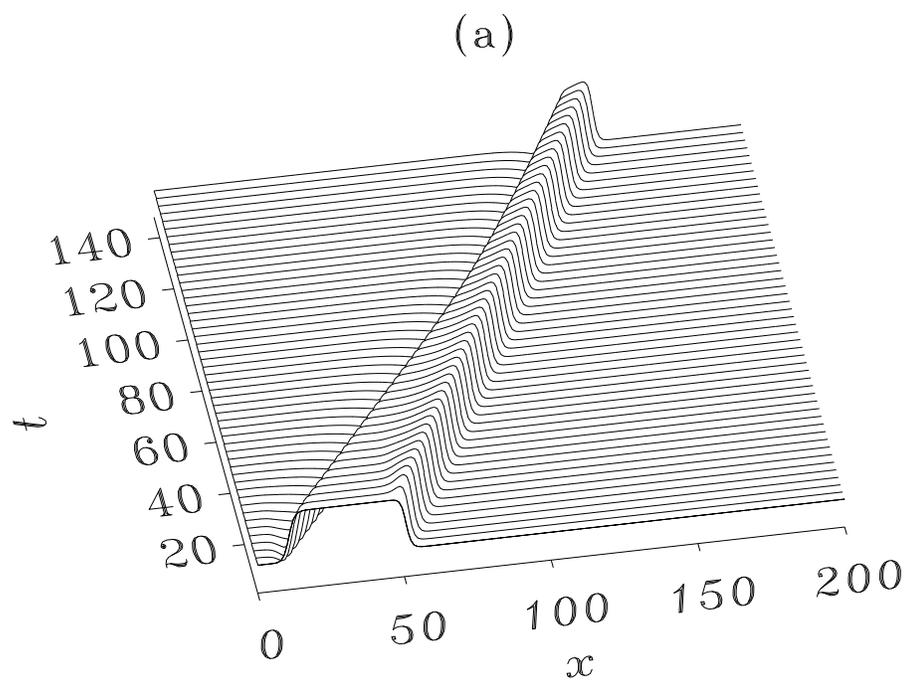
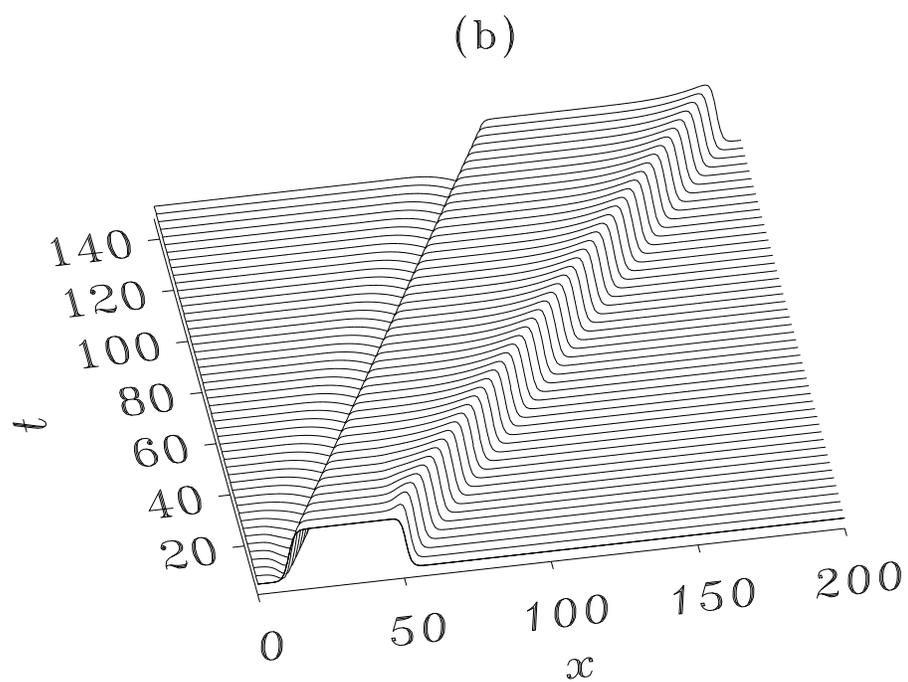
Time evolution of a front and a back (or two Bloch fronts following
one another). (a) The back propagates
faster than the front and binds to the front to form a traveling
up state domain. (b) The back propagates slower than the front and the up
state domain expands indefinitely.\hfill
\break
{\bf Figure 13:}~
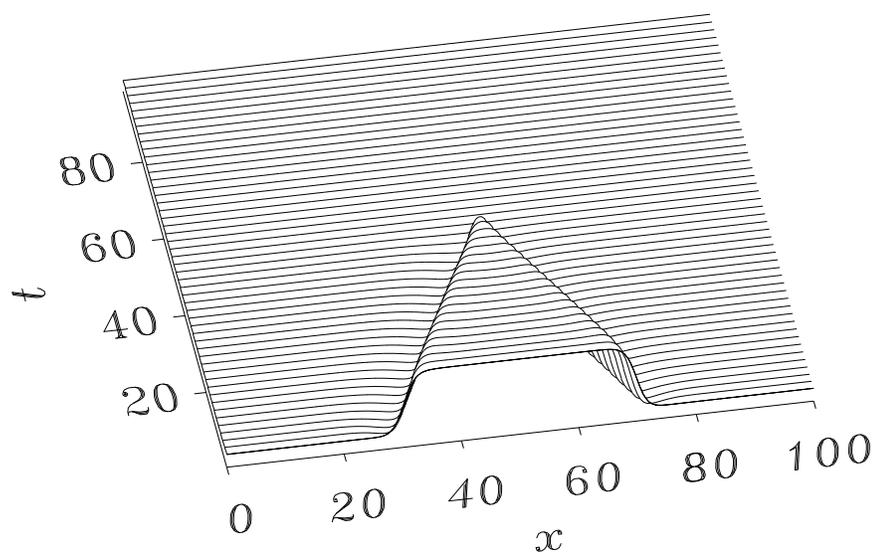
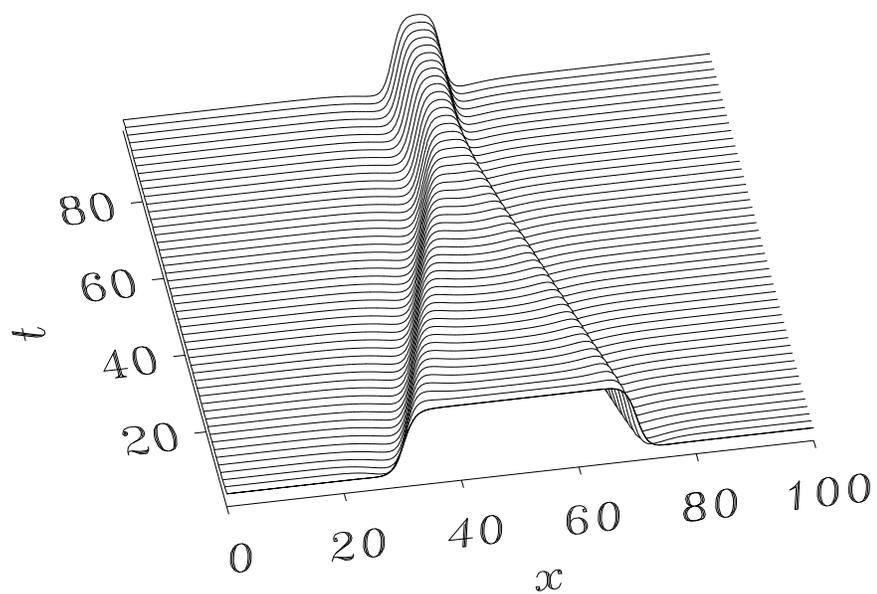
The onset of stationary domains in the Ising front regime as
$\delta/\epsilon$ is increased. Ising fronts either (a) collide and anihilate
to form a uniform state for $\epsilon>\mu_{st}\delta$, or (b) slow
to a stop and form a stationary domain for $\epsilon<\mu_{st}\delta$.\hfill
\break
{\bf Figure 14:}~
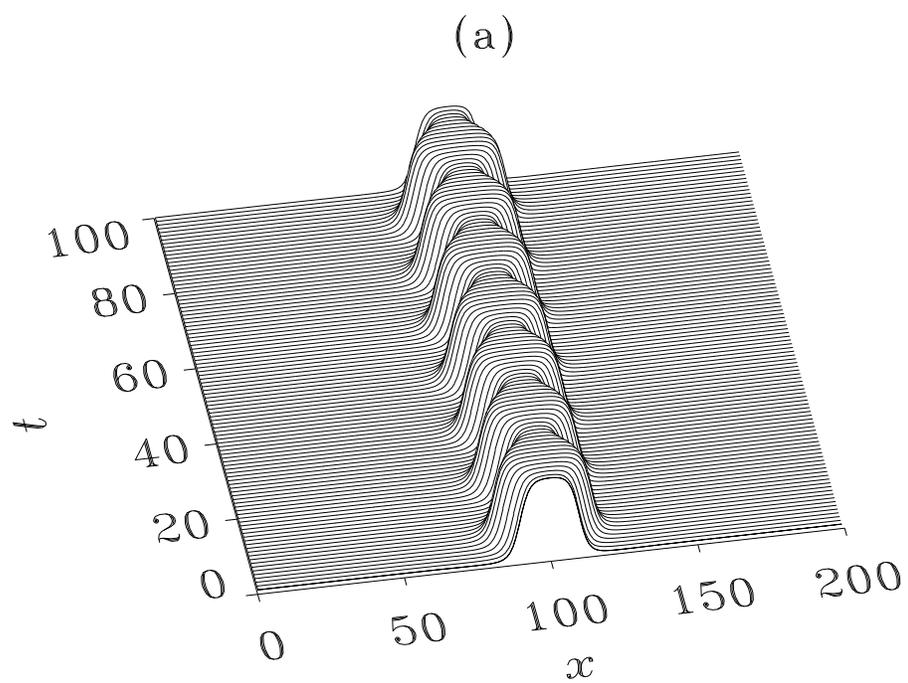
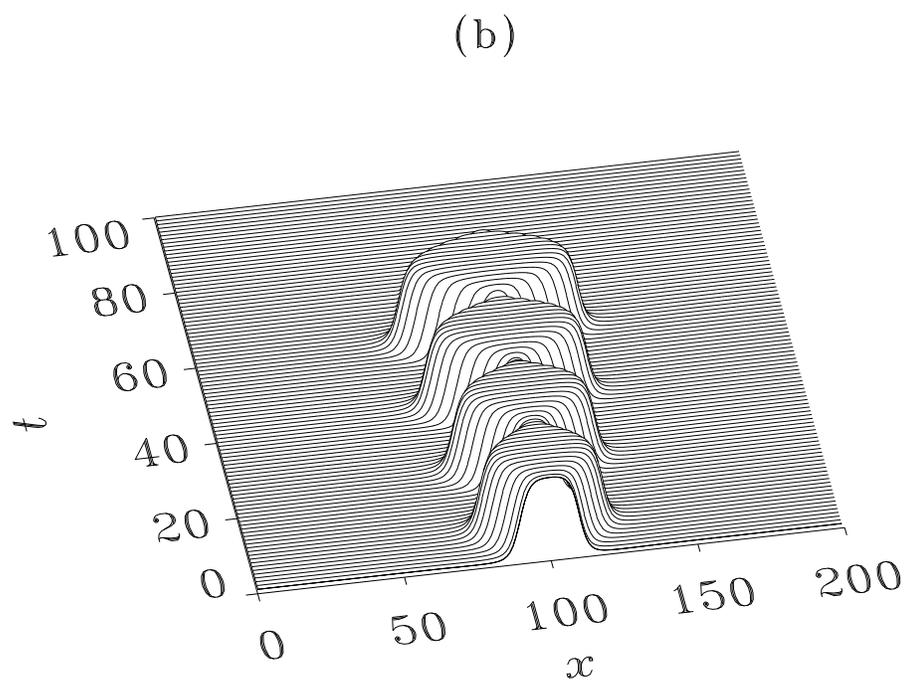
Oscillating or breathing domains.
(a) Steady oscillations close to the Hopf bifurcation.  Computational
parameters $\epsilon=0.03,\delta=2.5, a_1=2.0, a_0=-0.1$.
(b) Collapse of an oscillating domain further away from the Hopf bifurcation.
Compuational parameters are the same with $\epsilon=0.025$.\hfill
\break
{\bf Figure 15:}~
The effect of the front bifurcation on the dynammics of two fronts
propagating toward one another: (a) Below the bifurcation
($\epsilon>\epsilon_c(\delta)$ an oscillating domain is formed.
Computational parameters: $\epsilon=0.03,\delta=2.5, a_1=2.0, a_0=-0.01.$
(b)  Beyond the bifurcation the two fronts rebound from one another and
propagate to the boundaries. Same parameters as in (a) except that
$\epsilon=0.012$.\hfill
\break
{\bf Figure 16:}~
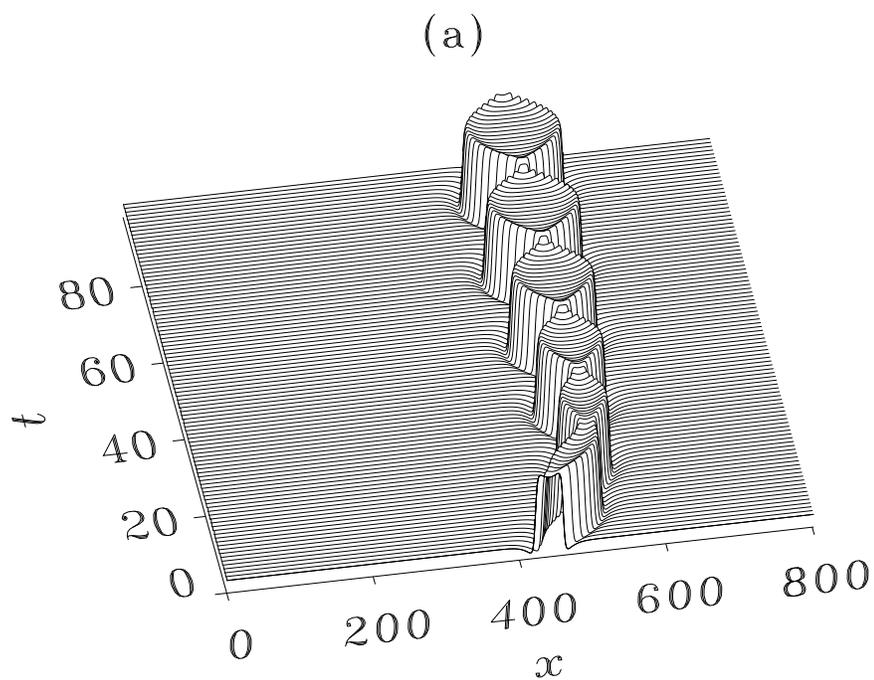
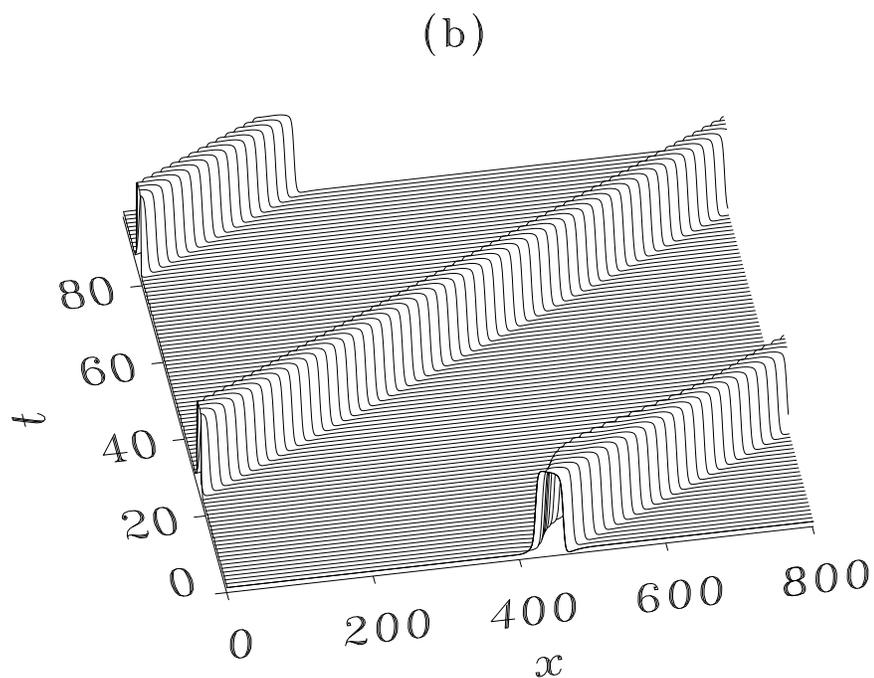
The effect of the front bifurcation on the dynamics of two fronts
following one another: (a) Below the bifurcation
($\epsilon>\epsilon_c(\delta))$ an oscillating domain is formed.
Computational parameters are $\epsilon=0.030,\delta=2.5, a_1=2.0, a_0=-0.012$.
(b) Beyond the bifurcation a traveling domain is formed. Same parameters as
in (a) except that $\epsilon=0.25$.\hfill
\break
{\bf Figure 17:}~
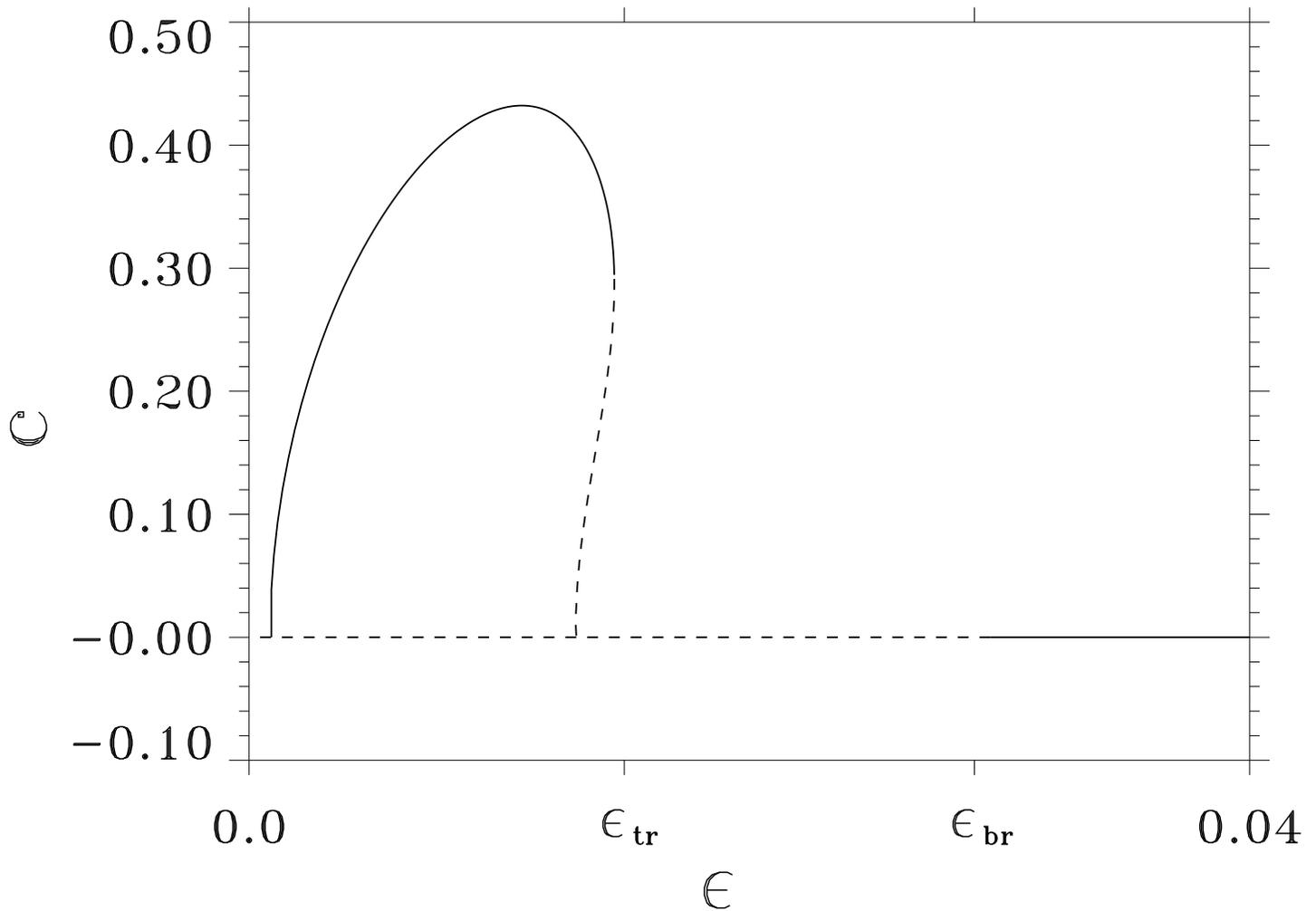
A typical bifurcation diagram for single domain structures.  The
solutions shown with the solid (dashed) line are stable (unstable) structures.
The stationary domain solution loses stability to an oscillating domain
at $\epsilon=\epsilon_{br}$. At $\epsilon=\epsilon_{tr}$, a branch
of stable traveling domain solutions appears. By the symmetry
$c \rightarrow -c, ~\chi \rightarrow -\chi$, this diagram is
symmetric with respect to the $c=0$ axis, but only the positive speed
branch of traveling domain solutions is shown. Also, there is an additional
branch of zero speed solutions that is unstable for all values of
$\epsilon$.  Parameters: $\delta=2.5,a_1=2.0,a_0=-.1$, period $=100$.\hfill
\break
{\bf Figure 18:}~
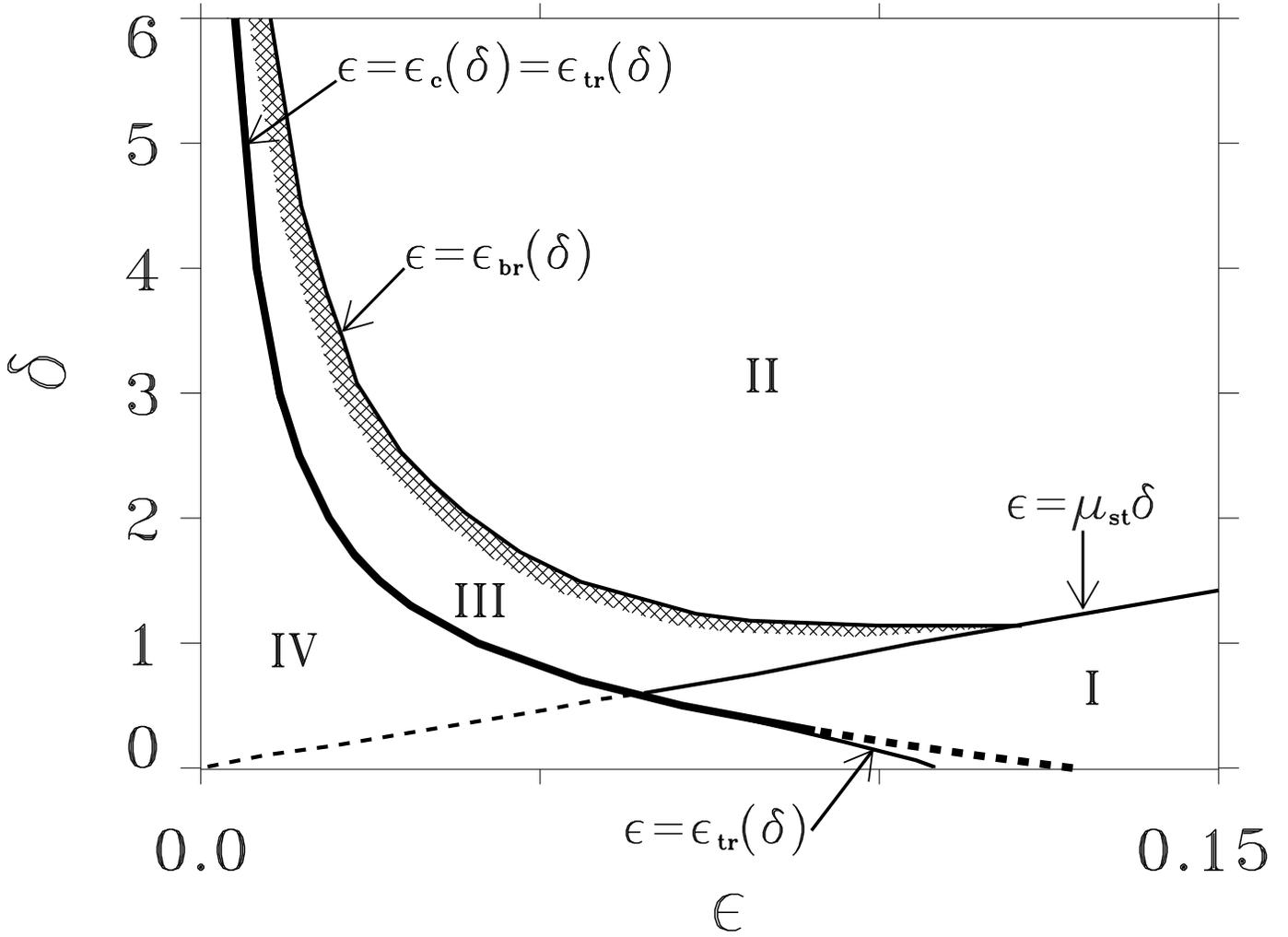
Phase diagram for single domain structures in the $\epsilon - \delta$
plane.  In region I, domain
structures are transient and the asymptotic state is either the uniform up
or down state. In region II stable stationary domains coexist with the
uniform up and down states.  In region
III domains oscillate with a typical region of steady oscillations denoted
by the hatched area.  In region IV domains travel. The boundary between
regions III and IV (i.e. the onset of traveling domains) coincides with the
front bifurcation line denoted by the thick solid/dashed curve.
For this phase diagram, $a_1=2.0$ and $a_0=-0.1$.\hfill
\break
{\bf Figure 19:}~
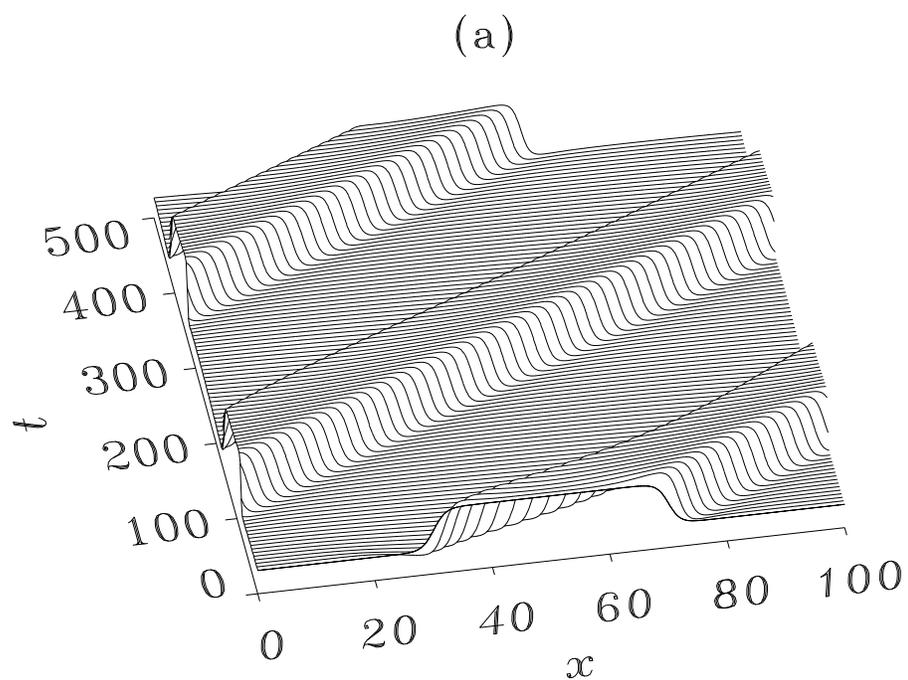
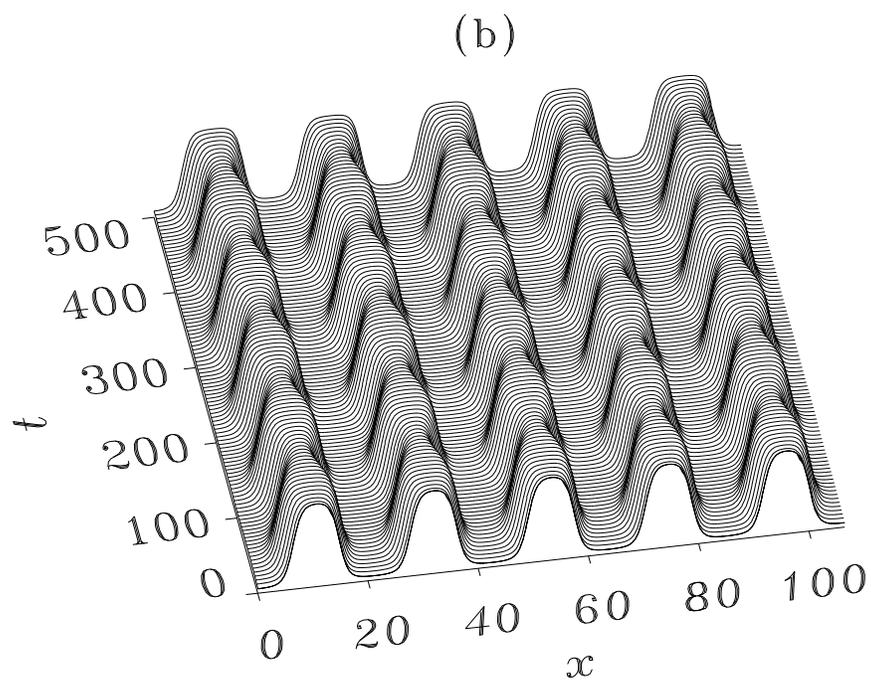
Coexistence of traveling and oscillating {\it periodic} patterns
beyond the bifurcation from Ising to Bloch fronts.
Both (a) and (b) are obtained for the same
computational parameters with different initial conditions.  Note the
difference in average wavelength between the two patterns. Computational
parameters: $\epsilon=0.013,\delta=2.5, a_1=2.0, a_0=-0.1$.\hfill
\break
{\bf Figure 20:}~
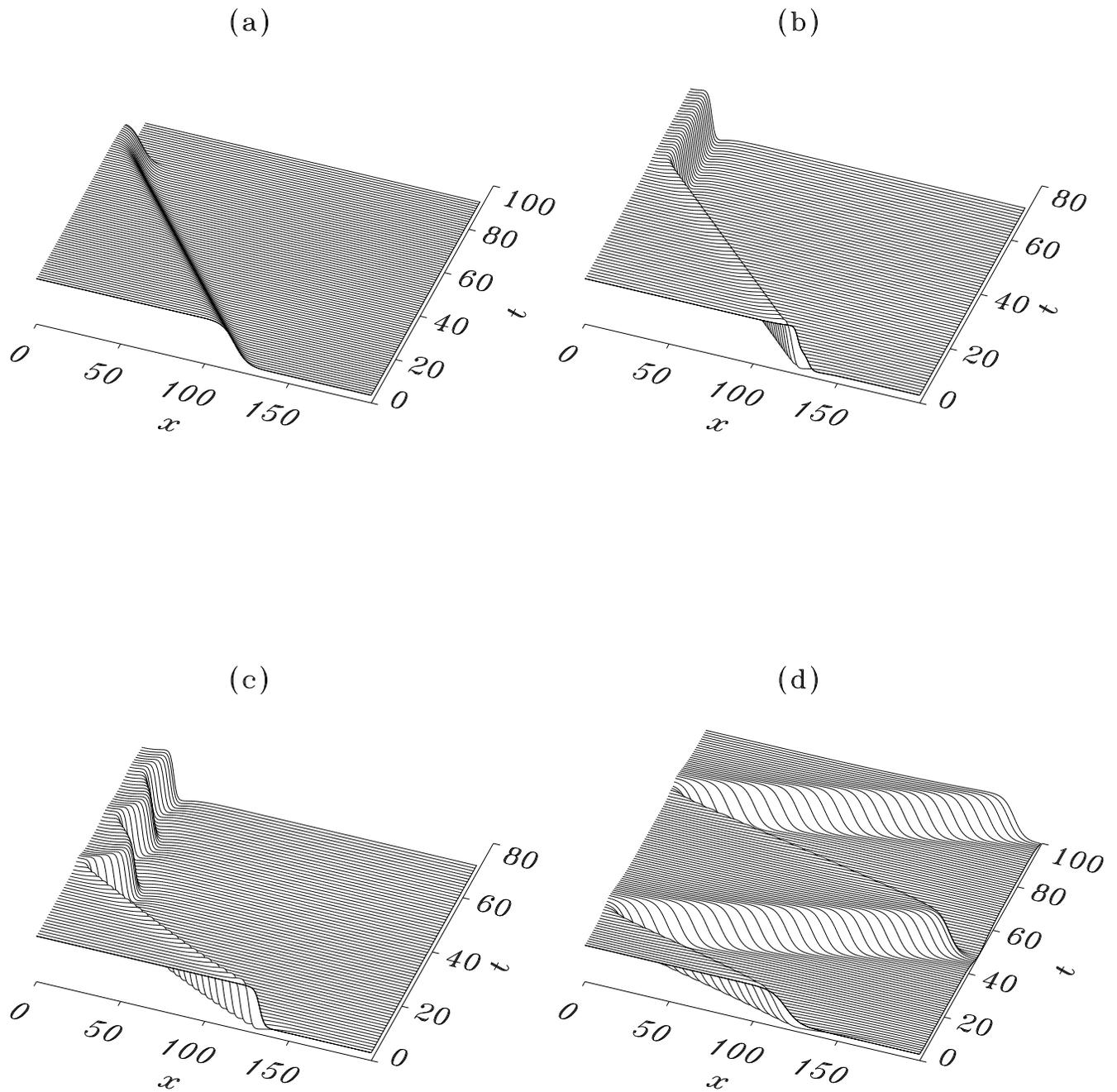
Front-boundary interactions. (a) Transient domain region: the
front is absorbed at the boundary.  (b)  Stationary domain region: the front
is stopped at the boundary.  (c) Oscillating domain region: the front
oscillates near the boundary. (d) Traveling domain region (beyond the
front bifurcation): the front is
reflected at the boundaries.  All simulations are with Neumann boundary
conditions.\hfill
\break

\vfil\bye